\documentclass{article}

\usepackage{arxiv}

\usepackage[utf8]{inputenc}
\usepackage[T1]{fontenc}
\usepackage{hyperref}
\usepackage{url}
\usepackage{booktabs}
\usepackage{subcaption}
\usepackage{amsfonts}
\usepackage{nicefrac}
\usepackage{microtype}
\usepackage{gensymb}
\usepackage{amsmath}
\usepackage{cleveref}
\usepackage{lipsum}
\usepackage{graphicx}
\usepackage{natbib}
\usepackage{doi}
\usepackage{geometry}
\usepackage[group-separator={,}]{siunitx}

\title{AIFS-CRPS: Ensemble forecasting using a model trained with a loss function based on the Continuous Ranked Probability Score}

\author{
    Simon Lang
    \And Mihai Alexe
    \And Mariana C. A. Clare
    \And Christopher Roberts
    \And Rilwan Adewoyin
    \And Zied {Ben Bouall\`egue}
    \And Matthew Chantry
    \And Jesper Dramsch
    \And Peter D. Dueben
    \And Sara Hahner
    \And Pedro Maciel
    \And Ana Prieto-Nemesio
    \And Cathal O'Brien
    \And Florian Pinault
    \And Jan Polster
    \And Baudouin Raoult
    \And Steffen Tietsche
    \And Martin Leutbecher
    \And European Centre for Medium-Range Weather Forecasts (ECMWF)
}

\renewcommand{\shorttitle}{AIFS-CRPS}

\begin{document}
\maketitle

\begin{abstract}
Over the last three decades, ensemble forecasts have become an integral part of forecasting the weather. They provide users with more complete information than single forecasts as they permit to estimate the probability of weather events by representing the sources of uncertainties and  accounting for the day-to-day variability of error growth in the atmosphere. This paper presents a novel approach to obtain a weather forecast model for ensemble forecasting with machine-learning.
 
AIFS-CRPS is a variant of the Artificial Intelligence Forecasting System (AIFS) developed at ECMWF. Its loss function is based on a proper score, the Continuous Ranked Probability Score (CRPS). For the loss, the almost fair CRPS is introduced because it approximately removes the bias in the score due to finite ensemble size yet avoids a degeneracy of the fair CRPS. 
The trained model is stochastic and can generate as many exchangeable members as desired and computationally feasible in inference.

For medium-range forecasts AIFS-CRPS outperforms the physics-based Integrated Forecasting System (IFS) ensemble for the majority of variables and lead times. For subseasonal forecasts, AIFS-CRPS outperforms the IFS ensemble before calibration and is competitive with the IFS ensemble when forecasts are evaluated as anomalies to remove the influence of model biases.
\end{abstract}

\section{Introduction}
Over the last few years, several machine-learned weather prediction models have emerged that show superior performance to that of traditional physics-based models in many forecast scores. The first generation of these models produced deterministic predictions and were trained to minimise a mean-squared-error (MSE) loss \citep{pathak2022fourcastnet,pangu2023,lam2022graphcast,Chen2023,lang2024aifs}. The MSE training objective incentivises the smoothing of forecast fields with lead-time and a reduction of forecast activity, to avoid the 'double-penalty' incurred when forecasting misplaced structures (e.g. \cite{Hoffman1995, ebertdouble}). Nevertheless, evaluations demonstrated that these models display surprisingly physical behaviour in classical forecast situations \citep{hakim2024dynamical}, and provide genuine, useful predictions, including of many extreme events \citep{benbouallegue2023rise}. The European Centre for Medium-Range Weather Forecasts (ECMWF) is now producing semi-operational weather predictions using the Artificial Intelligence Forecasting System (AIFS, \cite{lang2024aifs}) four times per day.

For the usefulness of a weather forecast it is important to account for forecast uncertainties. Ensemble forecasts are run to estimate the probability density of the atmospheric state at a future time \citep{lewis2005,leutbecher2008ensemble}. In physics-based numerical weather prediction (NWP), this is achieved via running the forecast model from a range of perturbed initial conditions and by introducing stochastic perturbations into the forecast model itself. The aim is to generate a well calibrated ensemble. This means that on average, the ensemble standard deviation needs to match the root-mean square error of the ensemble mean (e.g., \cite{fortin2014}), and the predicted probability of an event should accurately reflect the observed probability of it occurring.

For physics-based ensemble simulations with the Integrated Forecasting System (IFS) at ECMWF \citep{molteni1996ecmwf}, initial condition uncertainty is represented via an ensemble of data assimilations \citep{doi:10.1002/qj.346, isaksen2010eda, lang2019eda} and singular vector perturbations \citep{leutbecher2008ensemble}. Perturbations to the initial conditions of the ensemble members are then constructed from both (see \cite{lang2021more} for an up-to-date description of the initial perturbation methodology). Uncertainties associated with the forecast model are represented stochastically \citep{leutbecher2017, berner.ea.2017}.

For the first generation of machine-learned weather prediction models, ensemble forecasts were mainly based on an ensemble of MSE trained forecast models \citep{bihlo2021generative,scher2021ensemble, clare2021combining, pathak2022fourcastnet, pangu2023, weyn2024ensemble}. The resulting ensemble forecasts tend to have too little ensemble spread. This is in line with the models having difficulties to represent the inherent variability of the atmosphere due to their training regime \citep{blogaifs2}.

The second generation of machine-learned weather forecast models are based on probabilistic training. For example, \cite{kochkov2024neural} developed a hybrid model that combined a differentiable solver for atmospheric dynamics with a machine-learned physics module; this approach achieved good ensemble scores. Most notably, denoising diffusion \citep{DBLP:journals/corr/Sohl-DicksteinW15, karras2022elucidating} has been used successfully to create machine-learned ensemble models that are competitive with physics-based NWP models across a range of probabilistic forecast scores \citep{price2023gencast,ensemble_blog,aifsensemblesNL2024}. Next to providing useful information about forecast uncertainties, these models have more stable statistics than the deterministically trained models and do not smooth out small-scale structures with forecast lead time. This can make forecasts more stable, and in some cases allows for consistent simulations from months \citep{aifsensemblesNL2024} to many years \citep{kochkov2024neural}.

At ECMWF, the first approach towards a machine-learned ensemble model is also based on a diffusion approach \citep{ensemble_blog, aifsensemblesNL2024} that achieves competitive ensemble scores when compared to the physics-based 9~km IFS ensemble \citep{lang81380}. In this work, we take a different approach: we introduce AIFS-CRPS, a machine-learned ensemble forecast model that is based on optimising a probabilistic proper score objective, similar to \cite{pacchiardi2024probabilistic, Shokar_2024} and \cite{kochkov2024neural}. AIFS-CRPS learns how to represent model uncertainty, through shaping Gaussian noise. For training, we use the \textit{almost fair continuous ranked probability score (afCRPS)}, a modification to the \textit{fair continuous ranked probability score (fCRPS)} \citep{Ferro_etal2008,Ferro2013,leutbecher2019}. 

AIFS-CRPS is a highly skilful machine-learned ensemble forecast model that is competitive or superior to the physics-based IFS ensemble across forecast lead times ranging from days to subseasonal predictions. 

\section{Probabilistic training}\label{sec:prob_training}
The AIFS-CRPS architecture largely follows that of the deterministic AIFS v0.2.1 \citep{lang2024aifs} with the encoder-processor-decoder design. The encoder and decoder of AIFS-CRPS are transformer-based graph neural networks (GNNs), while the processor backbone is a sliding window transformer. In contrast to the deterministic AIFS, however, the training of the AIFS-CRPS model is inherently probabilistic. AIFS-CRPS uses 16 processor layers and an embedding dimension of 1024 with 8 attention heads. This results in 229 million parameters in total.

AIFS and AIFS-CRPS operate on reduced Gaussian grids, such as the octahedral reduced Gaussian grid \citep{Wedi2014}.
Depending on the input resolution, the processor grid is an O48 (for data on an O96 input grid) or O96 (for data on a N320 input grid) 
octahedral reduced Gaussian grid (see table \ref{table:grids} for more details on grids).

\begin{table}[htbp]
\centering
\begin{tabular}{|p{2cm}|p{2cm}|p{5cm}|p{4cm}|}
\hline
\textbf{Grid name} & \textbf{\# grid points} & \textbf{Approx. resolution in degrees} & \textbf{Approx. resolution in km} \\ 
\hline
O1280 & \num{6599580} & 0.1 & 9 \\
\hline
N320 & \num{542080} & 0.25 & 30 \\
\hline
O96 & \num{40320} & 1.0 & 110 \\
\hline
O48 & \num{10944} & 2.0 & 210 \\
\hline
O32 & \num{5248} & 3.0 & 310 \\
\hline
\end{tabular}
\caption{Description of the horizontal reduced Gaussian grids featured in this paper. Octahedral reduced Gaussian grids ("O" grids) are used throughout, with the exception of the N320, which is the native ERA5 resolution.}
\label{table:grids}
\end{table}

\begin{figure}[htpb]
\centering
\includegraphics[width=0.9\linewidth]{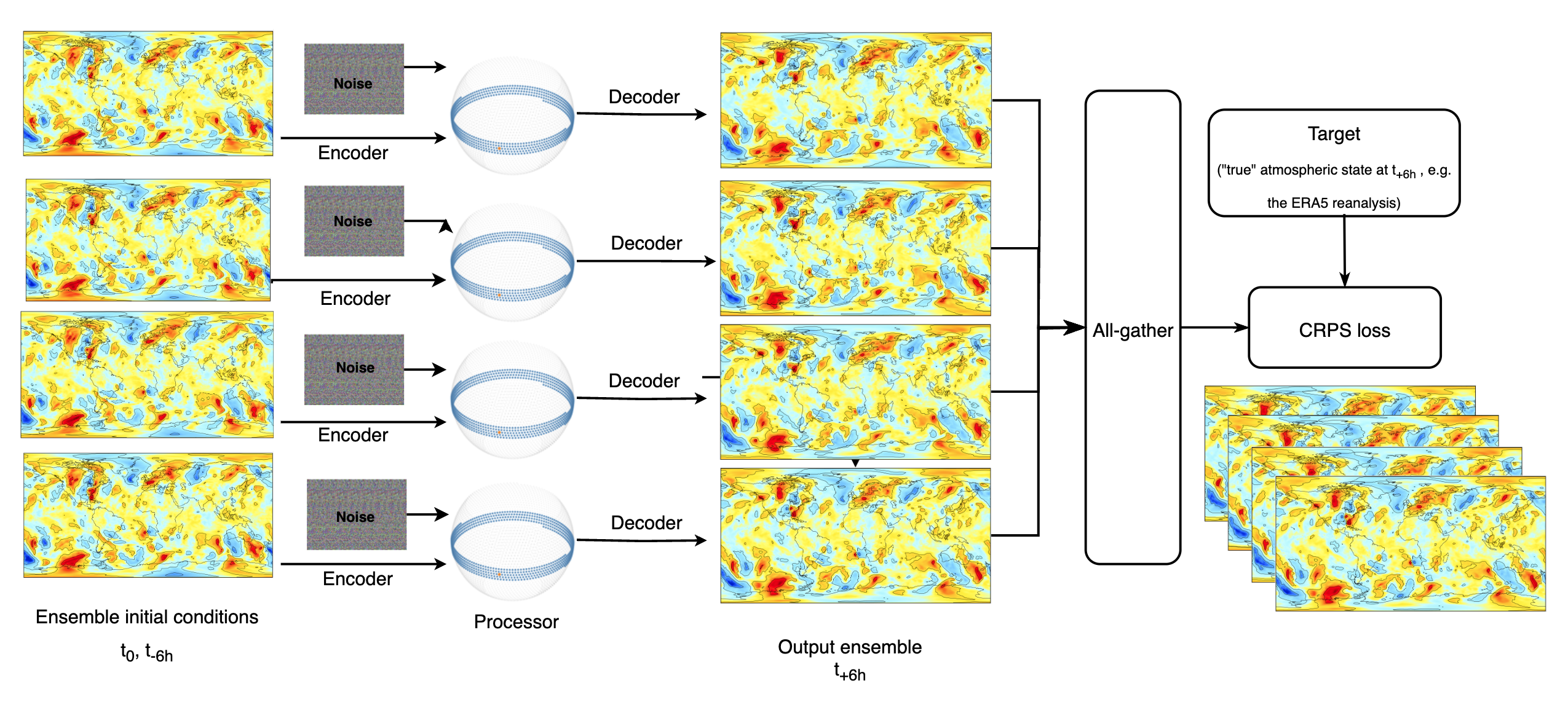}
\caption{Probabilistic training of AIFS-CRPS. A small ensemble of atmospheric states is propagated forward in time using separate model instances (that share the same weights). With ensemble sharding (see section \ref{sec:parallelism}), the ensemble forecasts are then gathered across all participating GPU devices using a differentiable all-gather operation. Finally, the (almost fair) CRPS loss is calculated from the AIFS-CRPS forecast ensemble and a deterministic analysis (e.g., ERA5) target.
}
\label{fig:aifs-crps-flow}
\end{figure}

For each forecast date, a small ensemble of states is propagated forward in time via independent model instances. These instances can either be initialised by the same atmospheric state (e.g. the ERA5 deterministic analysis) or from different initial conditions valid for the same date and time (e.g. generated from the ERA5 ensemble of initial conditions).
Here, we always initialise the ensemble members from the same ERA5 deterministic analysis during training.
For each model instance $i$ and forecast step, we generate an independent Gaussian noise sample ${\xi_i} \sim \mathcal{N}(0, \mathbf{I}_n)$, with $n$ equal to the number of processor grid points times the number of noise channels. Each random noise tensor is processed by a two-layer perceptron (MLP) followed by a layer normalisation operation \citep{ba2016layernormalization}. The noise embeddings are then used in conditional layer normalisations \citep{chen2021adaspeechadaptivetextspeech}, which replace all standard layer normalisations in the processor pre-norm transformer layers. The different ensemble members from the model instances are gathered and used to compute the probabilistic afCRPS loss (see section~\ref{sec:loss}). The loss is then minimised via backpropagation. It is important to note that the "ground truth" used here is deterministic, for example, the ERA5 deterministic analysis.
A schematic of this training set-up is depicted in figure~\ref{fig:aifs-crps-flow}. The forecast step can be chosen as required; here we set it to 6~h.

During inference mode, each ensemble member is initialised with a different random seed. The CRPS-trained ensemble members are completely independent and inference can run for each member in parallel. Given suitable initial conditions, it is possible to create as many ensemble members as required. 

When producing a forecast, the model is run in an auto-regressive fashion: the model is initialised from its own predictions, referred to as rollout. To improve forecast scores, rollout is also used in training \citep{keisler2022forecasting}, where the model learns to produce forecasts up to, e.g., 72~h into the future. Gradients flow through the entire forecast chain during backpropagation.

In the case of physics-based models, the perturbed ensemble members are usually constructed by introducing random numbers into the reference forecast model \citep{leutbecher2017}. Hence, in the AIFS-CRPS ensemble, there is no direct correspondence to an unperturbed (control) member often found in physics-based NWP ensemble systems, like the ECMWF ensemble \citep{moltenieps}, because the training  process is inherently probabilistic.

\subsection{Mitigating error accumulation during rollout}\label{sec:erroraccum}
AIFS computes the output state as a combination of a reference state - the input state - and a forecast tendency. 
The forecast tendency is the difference between the output state and the reference state. While it allows the model to focus on the forecast tendency, it can lead to errors accumulating when the model forecasts multiple steps auto-regressively. The output state is now an accumulation of all forecast tendencies produced by the forecast model. To advect a weather feature, the model has to create a tendency that exactly cancels out the feature at its previous location. Slight errors result in artefacts being left behind, which then affect the next model time step. 
To mitigate this effect, we downsample the reference state to lower resolution and then upsample it back to the original resolution. Then the forecast tendency is added to form the output state that enters the loss calculation:
\begin{align}
    x_{t+1} = U(D(x_t)) + f(x_t),
\end{align}
with the upsampling and downsampling operators $U$, $D$ and the forecast model $f(x)$. This still allows the model to focus on the forecast tendency, without the need to re-create the full state at each step. The model has full control over the output states. At the same time, small scale features in the reference state from the previous time step are removed. For up and downsampling, we make use of interpolation matrices generated by ECMWF's Meteorological Interpolation and Regridding (MIR) software package \citep{maciel2017mir}. The interpolation operators can be seen as graph convolutions without learnable parameters and can be efficiently implemented via sparse matrix multiplications. We have chosen an O32 reduced Gaussian grid (approximately $2.5\degree$) for the downsampling.

\subsection{Loss functions}\label{sec:loss}
Given an $M$-member forecast ensemble with members $\left\{ x_j \right\}_{j = 1 \ldots M}$, and $y$ the verifying observation (or analysis), the continuous ranked probability score (CRPS; see, e.g., \cite{hersbach2000}) is defined as:

\begin{align}
\text{CRPS}(\{x_j\}_{j=1}^{M}, y) &= \frac{1}{M} \sum_{j=1}^{M} |x_j - y| - \frac{1}{2M^2} \sum_{j=1}^{M} \sum_{k=1}^{M} |x_j - x_k|.\tag{1}
\label{crps_original}
\end{align}

The \textit{fair} CRPS is a modification to (\ref{crps_original}) that adjusts for ensemble size, penalising ensembles whose members do not behave as if they and the verifying observation were sampled from the same distribution \citep{Ferro2013,leutbecher2019}:

\begin{align}
\text{fCRPS}(\{x_j\}_{j=1}^{M}, y) &= \frac{1}{M} \sum_{j=1}^{M} |x_j - y| - \frac{1}{2M(M-1)} \sum_{j=1}^{M} \sum_{k=1}^{M} |x_j - x_k|. \tag{2}
\end{align}

However, the fCRPS suffers from a degeneracy in the case where all members apart from one have the same value as the verifying observation. In this case the remaining member is unconstrained and can take any value without impacting the fCRPS value. For computational efficiency, machine-learned models are commonly trained using reduced precision - float16 or lower, as described in, e.g., \cite{micikevicius2018mixedprecisiontraining} - and score degeneracy can become more likely. To avoid issues with score degeneracy we introduce the \textit{almost fair} CRPS:

\begin{align}
\text{afCRPS}_{\alpha} &:= \alpha \, \text{fCRPS} + (1-\alpha) \text{CRPS} \nonumber \\
&= \frac{1}{M} \sum_{j=1}^{M} |x_j - y| - \frac{M - 1 + \alpha}{2M^2(M-1)} \sum_{j=1}^{M} \sum_{k=1}^{M} |x_j - x_k| \nonumber \\
&= \frac{1}{M} \sum_{j=1}^{M} |x_j - y| - \frac{1 - \epsilon}{2M(M-1)} \sum_{j=1}^{M} \sum_{k=1}^{M} |x_j - x_k| \tag{3}
\label{afcrps_original}
\end{align}

with $\epsilon := \frac{(1-\alpha)}{M}$. Here the level $\alpha \in (0, 1]$ (and hence $\epsilon$) are model hyperparameters. A level of $\alpha=1$ corresponds to the fair CRPS and the intention is to use values of $\alpha$ close to 1 in order to obtain an almost fair score. 

To avoid numerical stability issues with finite precision when using afCRPS as the training objective, we rearrange (\ref{afcrps_original}) as a summation of positive terms:
\begin{align}
\text{afCRPS}_{\alpha} &= \frac{1}{2M(M-1)} \sum_{j=1}^{M} \sum_{\substack{k=1 \\ k \neq j}}^{M} \left( |x_j - y| + |x_k - y| - (1-\epsilon) |x_j - x_k| \right) \tag{4}
\end{align}
Using the triangle inequality, one sees that each term in the double sum is non-negative for $\epsilon \geq 0$.

The scaling used on each variable in the loss function are largely unchanged from those used in AIFS. In addition to per-variable loss scaling, we use a pressure dependent weighting factor for upper-air variables: a linear scaling according to $w_{pl} = plev/1000$, and optionally, restrict the minimum pressure scaling factor to 0.2.

\subsection{Training}
We train AIFS-CRPS in four stages. The first training stage follows \cite{lam2022graphcast}. The model learns to forecast one 6~h step forward in time (rollout=1). In the second stage we train AIFS-CRPS auto-regressively for two 6~h forecast steps, i.e. rollout 2. During the third phase, AIFS-CRPS is trained for multiple rollout steps. Here, the maximum rollout window is incremented after a certain number of epochs, increasing from 3 to 12 (18~h to 72~h), the learning rate is held constant during this phase. The final fourth stage is a fine-tuning phase, where the model is trained on the operational IFS analysis, going through a full rollout training, again up to step 12. The first training phase comprises a total of \num{300000} iterations (parameter updates), with an initial learning rate of $1e-3$. We use a cosine schedule with 1000 warm-up steps, during which the learning rate increases linearly from zero to the initial learning rate. The learning rate is then reduced from its maximum value to zero. The second phase consists of \num{60000} iterations and the third phase of approximately \num{45000} iterations. In the second stage we again use a cosine schedule, now with 100 warm-up steps and a initial learning rate of $1e-5$. We use a batch size of 16.

The loss is afCRPS with $\alpha=0.95$ and AdamW (\cite{loshchilov2018decoupled}) is used as the optimizer with $\beta$-coefficients set to 0.9 and 0.95, and a weight decay setting of 0.1. We concatenate the identifier of the time step to the initial state before embedding: 1 for the first forecast step, where the model is initialised from an analysis state, and 2 for subsequent auto-regressive forecast steps.

\subsection{Parallelism and memory management}\label{sec:parallelism}
The set-up of our model means we can parallelise the training of the model in three different and complementary ways. The simplest option is to perform data parallelism \citep{li2020ddp}, where a batch is divided into smaller sub-batches and these are processed simultaneously across multiple GPUs. 

However, to allow for larger ensemble sizes, bigger models and longer rollouts, AIFS also supports two different sharding levels (\cite{lang2024aifs}). Model instances as well as ensemble groups can be split across several GPUs, e.g., an ensemble group consisting of four ensemble members can be split across 16 GPUs, with four GPUs per ensemble member. This makes it possible to train models with larger parameter counts and higher spatial resolution. Model sharding is required when training an ensemble model at higher spatial resolution. For example when using N320 input data, the native resolution of the ERA5 reanalysis.

Like AIFS, the AIFS-CRPS makes extensive use of activation checkpointing \citep{chen2016training} during the forward pass of the model; this includes the afCRPS loss calculation.

\section{Datasets}
In training, we use the Copernicus ERA5 reanalysis dataset produced by ECMWF \citep{hersbach2020era5} for both the initial conditions and the afCRPS objective. For fine-tuning we also use the operational IFS analysis. As input, we provide a representation of the atmospheric states at $t_{-6h}$, $t_0$ to forecast the state at time $t_{+6h}$, as is the case in AIFS and many other machine-learned weather prediction models. We use the years 1979 to 2017 for training. For fine-tuning on the operational IFS analysis, we use the years 2016 to 2023.

During inference, the ensemble members start from the initial conditions of the operational ECMWF ensemble. These are constructed by combining perturbations from the ECMWF ensemble of data assimilations system with perturbations based on singular vectors (see \cite{lang2021more} for details).

The input and output fields of AIFS-CRPS are mostly similar to those of the AIFS and are shown Table~\ref{table:variables}. 
The analysis states are interpolated from their native resolution (ERA5: N320, \num{542080} grid points; Operational IFS analysis: TCo1279, \num{6599680} grid points) to the AIFS-CRPS input grid resolution for the forecast initialisation, when required. 
\begin{table}[htbp]
\centering
\begin{tabular}{|p{5cm}|p{3cm}|p{3cm}|p{3cm}|}
\hline
\textbf{Field / Variable} & \textbf{Level type} & \textbf{Input/Output} & \textbf{Normalisation} \\ 
\hline
Geopotential, horizontal and vertical wind components, specific humidity, temperature & Pressure level: 50, 100, 150, 200, 250, 300, 400, 500, 600, 700, 850, 925, 1000 & Both & {Standardised, apart from the geopotential, which is max-scaled} \\ 
\hline
Surface pressure, mean sea-level pressure, skin temperature, 2 m temperature, 2 m dewpoint temperature, 10 m horizontal wind components, total column water & Surface & Both & {Standardised} \\ 
\hline
Total precipitation & Surface & Output & {Standard deviation changed but mean kept the same}\\ 
\hline
Land-sea mask, orography, standard deviation and slope of sub-grid orography,  insolation, cosine and sine of latitude and longitude, cosine and sine of the local time of day and day of year & Surface & Input & {Sub-grid orography max-scaled; all others not normalised} \\
\hline
Time step identifier & - & Input & -\\
\hline
\end{tabular}
\caption{Input and output variables of AIFS-CRPS.}
\label{table:variables}
\end{table}

\section{Experiments}\label{sec:experiments}
We train two AIFS-CRPS versions with different grid configurations: The lower resolution version has an O96 input grid and an O48 processor grid (see table \ref{table:grids} for more detail on the grids). The higher resolution version has an N320 input grid and an O96 processor grid. 
Apart from input and processor resolution, the training set-up of the N320 model differs from the O96 set-up in two ways: we train with 2 ensemble members only to mitigate costs, and we do not yet use a minimum pressure scaling factor (see section~\ref{sec:loss}). However, this will be revised in the future. The different training settings of the experiments are summarised in Table~\ref{table:exps}.
\begin{table}[htbp]
    \centering
    \begin{tabular}{| m{2.8cm} | m{12cm} |}
        \hline
        \textbf{Experiment} & \textbf{Settings} \\ 
        \hline
        AIFS-CRPS O96 & 
        \begin{itemize}
            \item ensemble size 4
            \item linear pressure scaling with minimum 0.2
            \item rollout schedule, learning rate 1e-6, increment every epoch
            \item fine-tune rollout schedule, learning rate 5e-7, increment every other epoch
            \item trained for $\sim$ 4 days on 64 H100 64GB GPUs
        \end{itemize} \\
        \hline
        AIFS-CRPS N320 &
        \begin{itemize}
            \item ensemble size 2
            \item linear pressure scaling
            \item rollout schedule, lr 1e-6, increment every epoch
            \item fine-tune rollout schedule, lr 5e-7, increment every other epoch
            \item trained for $\sim$ 7 days on 128 H100 64GB GPUs
        \end{itemize} \\
        \hline
    \end{tabular}
    \caption{The training settings used for AIFS-CRPS O96 and N320.}
    \label{table:exps}
\end{table}

\section{Evaluation}
\subsection{Variability}\label{variability}
\begin{figure}[htpb]
\centering
\begin{subfigure}{0.45\linewidth}
    \includegraphics[width=\linewidth]{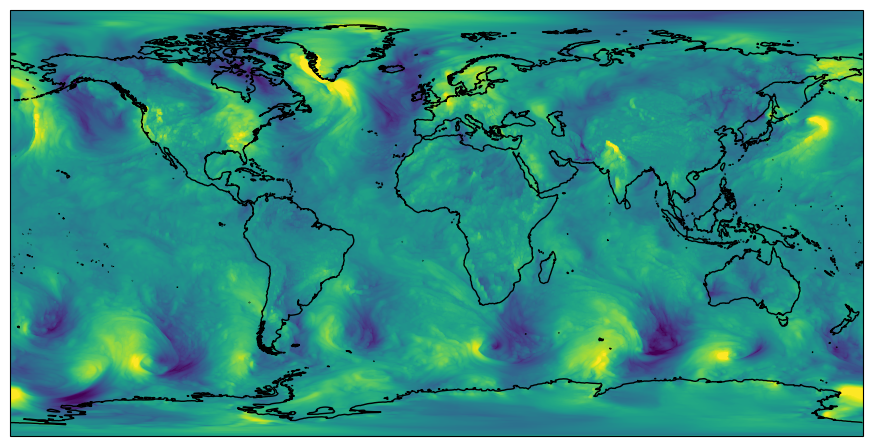}\caption{}\label{fig:24h_ifs}
\end{subfigure}
\hfill
\begin{subfigure}{0.45\linewidth}
    \includegraphics[width=\linewidth]{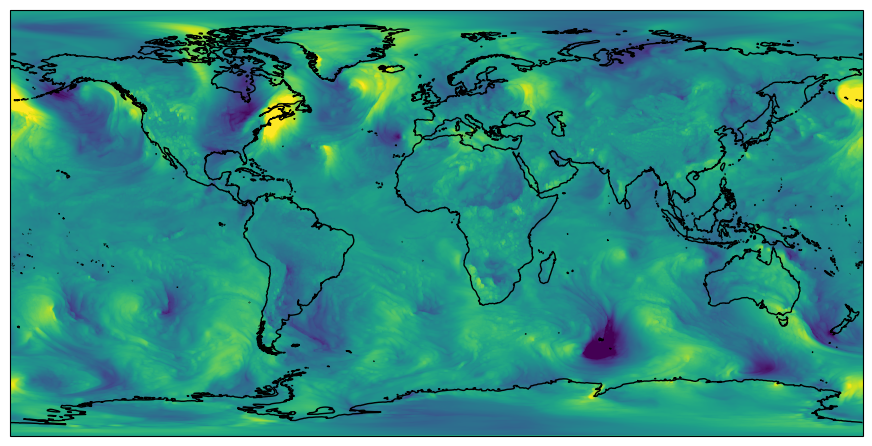}\caption{}\label{fig:240h_ifs}
\end{subfigure}
\begin{subfigure}{0.45\linewidth}
    \includegraphics[width=\linewidth]{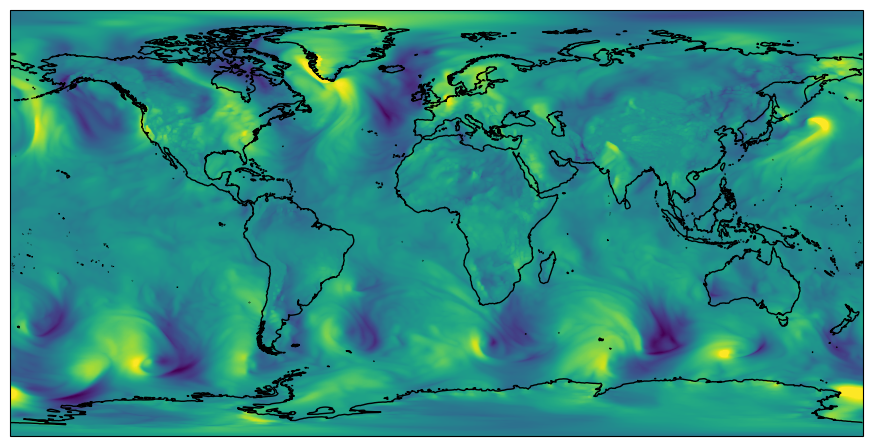}\caption{}\label{fig:24h_aifs}
\end{subfigure}
\hfill
\begin{subfigure}{0.45\linewidth}
    \includegraphics[width=\linewidth]{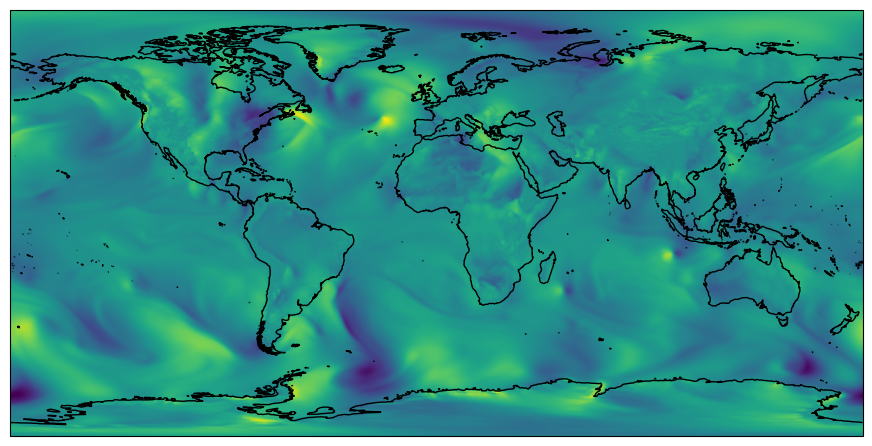}\caption{}\label{fig:240h_aifs}
\end{subfigure}
\begin{subfigure}{0.45\linewidth}
    \includegraphics[width=\linewidth]{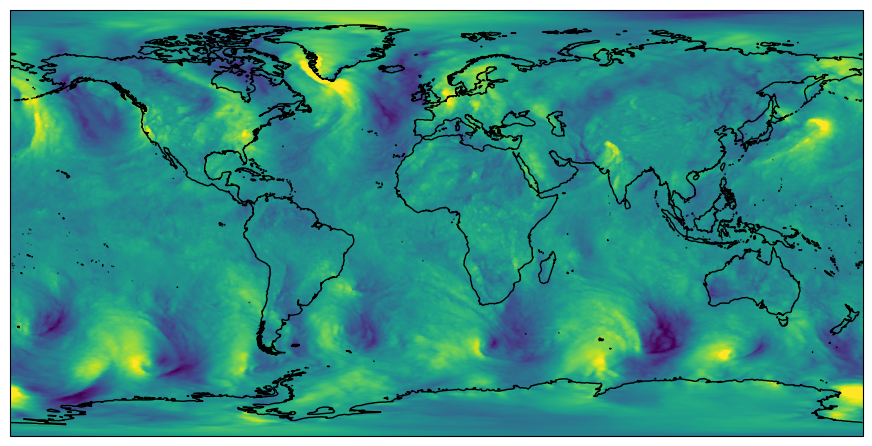}\caption{}\label{fig:24h_aifs-kcrps-n320}
\end{subfigure}
\hfill
\begin{subfigure}{0.45\linewidth}
    \includegraphics[width=\linewidth]{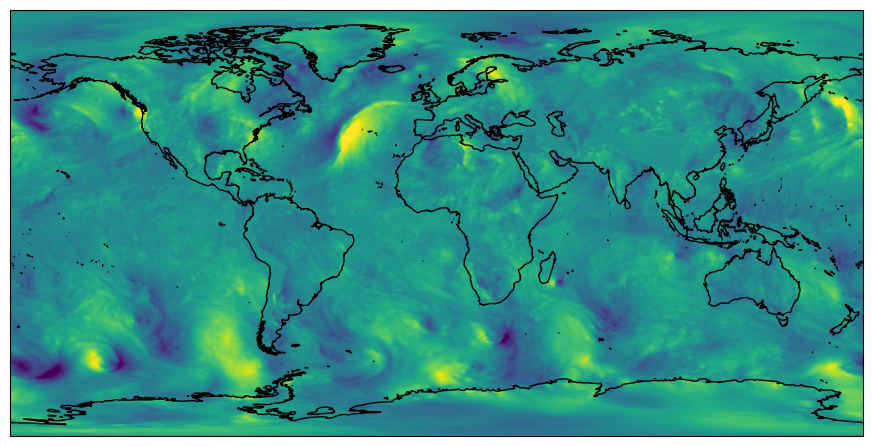}\caption{}\label{fig:240h_aifs-kcrps-n320}
\end{subfigure}
\begin{subfigure}{0.45\linewidth}
    \includegraphics[width=\linewidth]{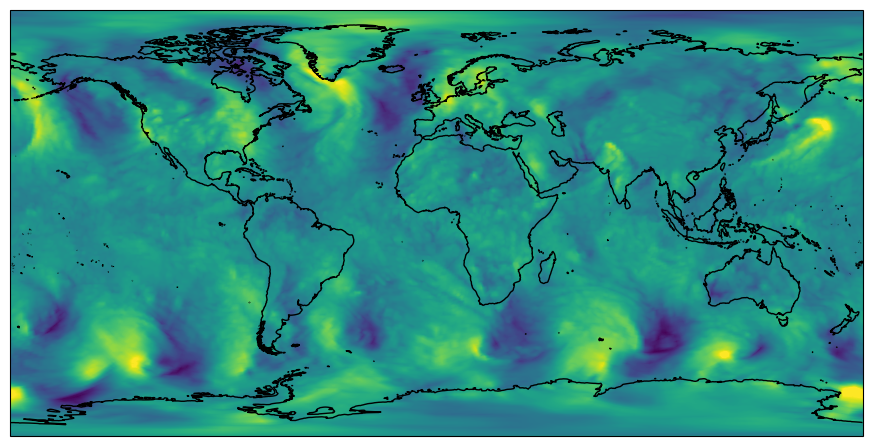}\caption{}\label{fig:24h_aifs-kcrps-o96}
\end{subfigure}
\hfill
\begin{subfigure}{0.45\linewidth}
    \includegraphics[width=\linewidth]{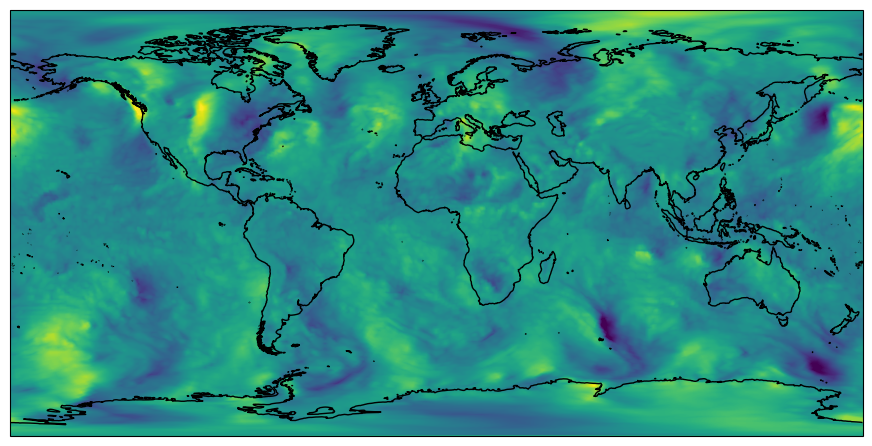}\caption{}\label{fig:240h_aifs-kcrps-o96}
\end{subfigure}
\caption{24-hr (left column) and 240-hr (right column) forecasts of meridional wind at 850~hPa, from perturbed member 1 of the IFS 9~km ensemble (approximately 0.1$\degree$ spatial resolution, \subref{fig:24h_ifs}, \subref{fig:240h_ifs}) AIFS trained with a MSE loss (approximately 0.25$\degree$ spatial resolution; \subref{fig:24h_aifs}, \subref{fig:240h_aifs}), perturbed member 1 of the AIFS-CRPS N320 ensemble (approximately 0.25$\degree$ spatial resolution; \subref{fig:24h_aifs-kcrps-n320}, \subref{fig:240h_aifs-kcrps-n320}) and of the AIFS-CRPS O96 ensemble (approximately 1.0$\degree$ spatial resolution; \subref{fig:24h_aifs-kcrps-o96}, \subref{fig:240h_aifs-kcrps-o96}). The forecasts are initialised on March 1st 2024, 00 UTC. For plotting, the fields have been interpolated to a regular 0.25$\degree$ latitude-longitude grid.}
\label{fig:v850}
\end{figure}

In contrast to IFS (figure~\ref{fig:24h_ifs} and \ref{fig:240h_ifs}), the AIFS trained with an MSE loss loses small-scale detail with forecast lead time (compare figure~\ref{fig:24h_aifs} and figure~\ref{fig:240h_aifs}). However, the ensemble members of the AIFS-CRPS ensemble maintain realistic variability throughout the forecast range, with no visible blurring (compare figure~\ref{fig:24h_aifs-kcrps-n320} and \ref{fig:240h_aifs-kcrps-n320}, and figure~\ref{fig:24h_aifs-kcrps-o96} and \ref{fig:240h_aifs-kcrps-o96}). This is an important property for ensemble forecasts and the representation of extreme events such as intense mesoscale and synoptic-scale weather systems, for example tropical cyclones and extra-tropical storms.

\begin{figure}[htpb]
    \centering
    \begin{subfigure}{0.40\linewidth}
        \includegraphics[width=\linewidth]{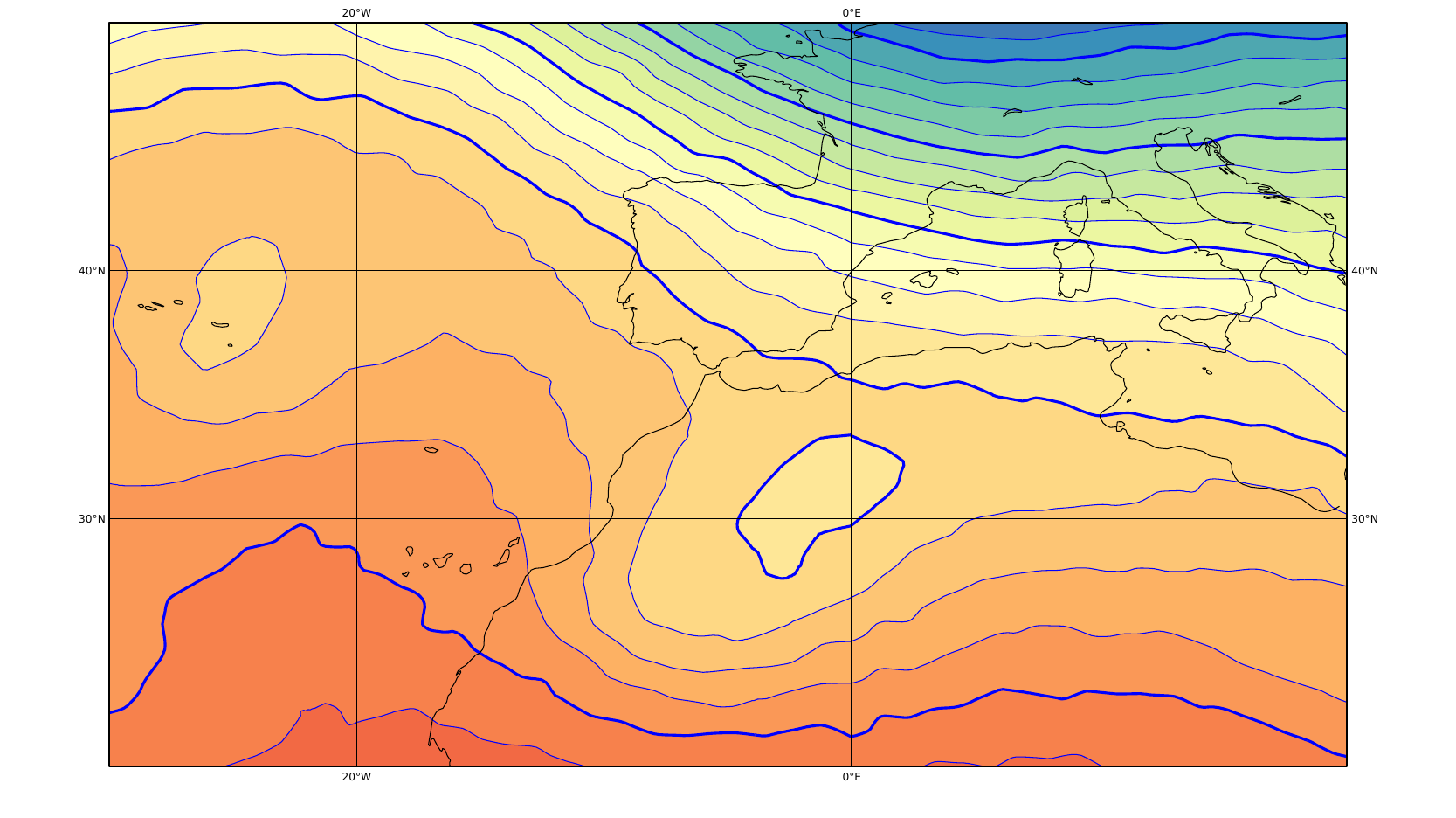}
        \caption{}\label{fig:noise1}
    \end{subfigure}
    \hspace{2.5cm}
    \begin{subfigure}{0.40\linewidth}
        \includegraphics[width=\linewidth]{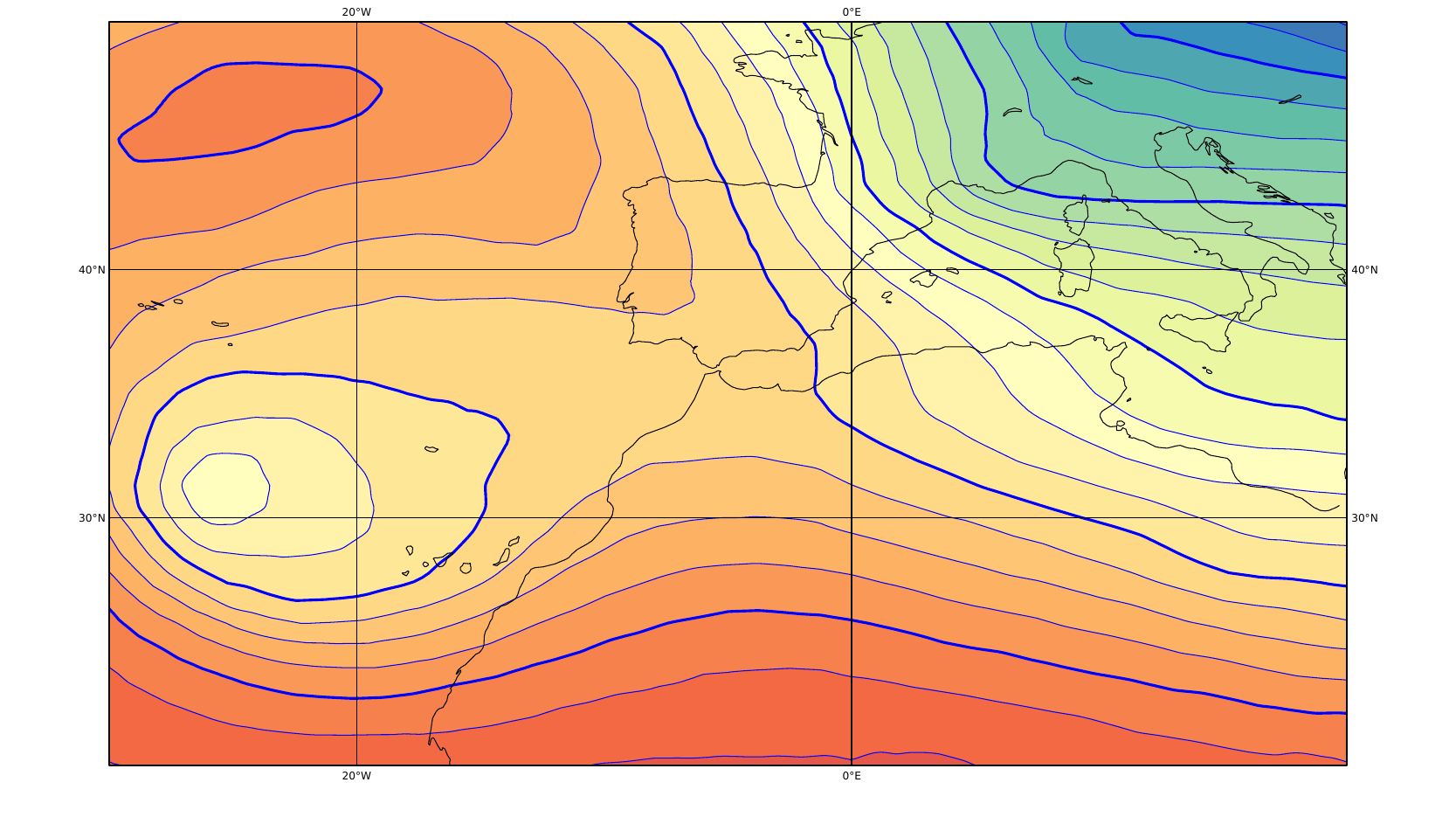}
        \caption{}\label{fig:noise2}
    \end{subfigure}
    \caption{Geopotential at 500~hPa of a 300-hour forecast from perturbed member~1 of the AIFS-CRPS ensemble when the model generates a tendency with respect to the full resolution input (reference) field (\subref{fig:noise1}) compared to when the tendency is generated with respect to a truncated input (reference) field (\subref{fig:noise2}). For more explanation, see section~\ref{sec:erroraccum}.}
    \label{fig:noise}
\end{figure}

First results without reference field truncation (see section~\ref{sec:erroraccum}) exhibited spurious  increase in variability with forecast lead time for  fields that tend to be smooth in the analysis such as geopotential at 500~hPa (figure~\ref{fig:noise}). It starts at small scale but then propagates to larger scales. The increase of variability is visible when comparing contour plots. Without reference field truncation, the contour lines become wiggly at longer lead times because the fields gain excess small-scale energy (figure~\ref{fig:noise1}). With reference field truncation, the contour lines appear smooth (figure~\ref{fig:noise2}, forecasts here have been run with different random seeds).

\begin{figure}[htpb]
\centering
\begin{subfigure}{0.45\linewidth}
    \includegraphics[width=\linewidth]{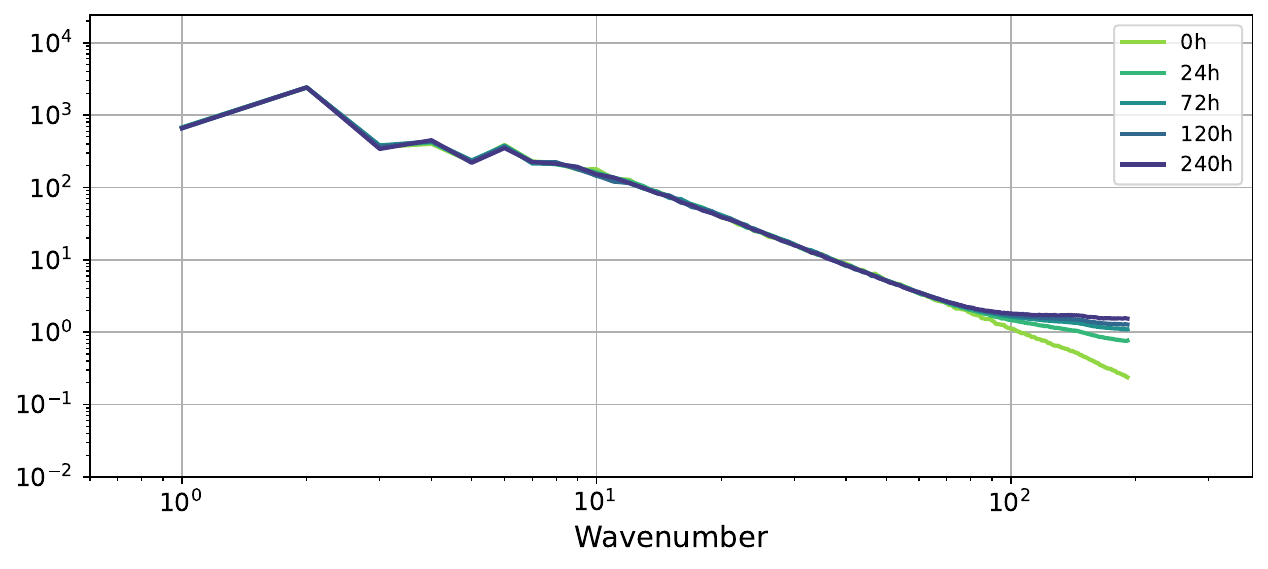}
    \caption{}\label{fig:spc_z_1}
\end{subfigure}
\hfill
\begin{subfigure}{0.45\linewidth}
    \includegraphics[width=\linewidth]{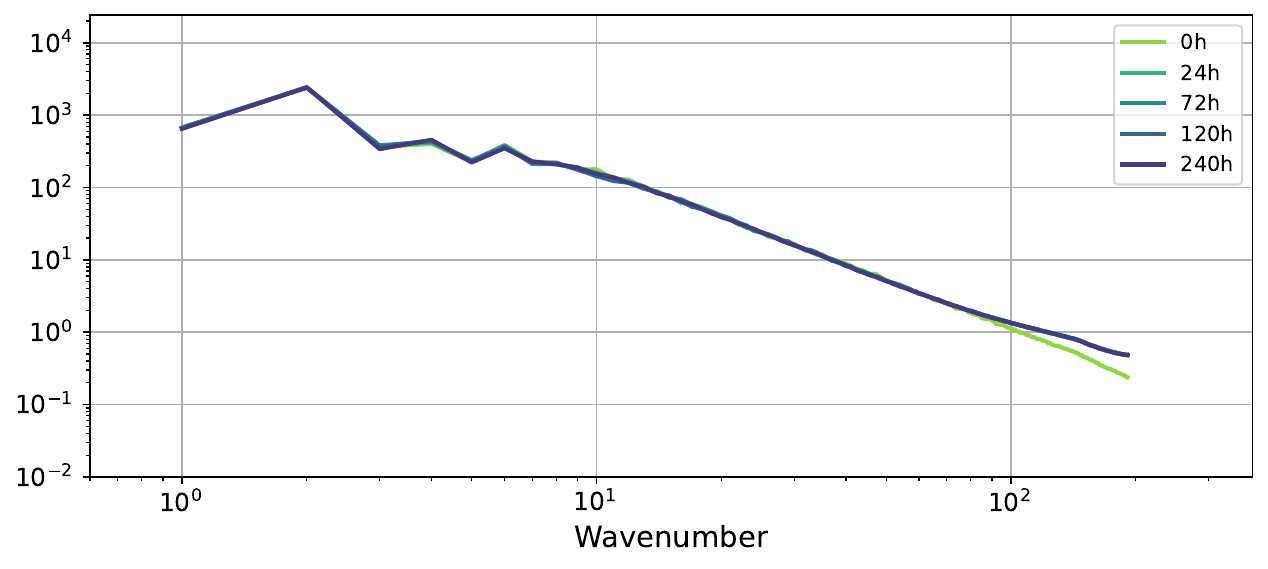}
    \caption{}\label{fig:spc_z_2}
\end{subfigure}
\begin{subfigure}{0.45\linewidth}
    \includegraphics[width=\linewidth]{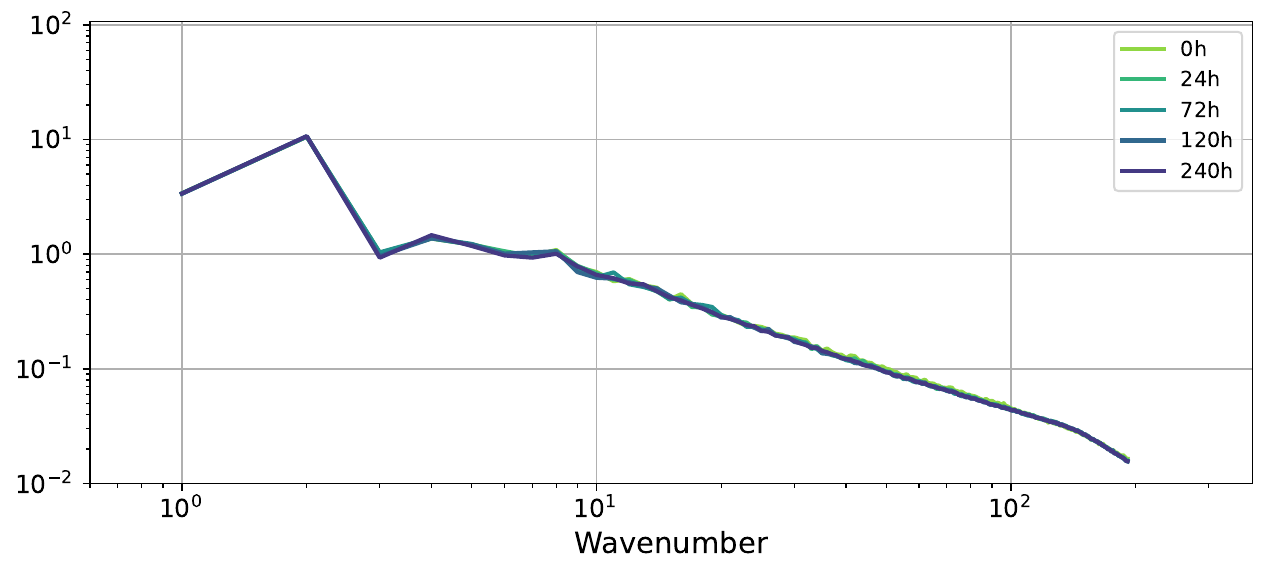}
    \caption{}\label{fig:spc_t_2}
\end{subfigure}
\hfill
\begin{subfigure}{0.45\linewidth}
    \includegraphics[width=\linewidth]{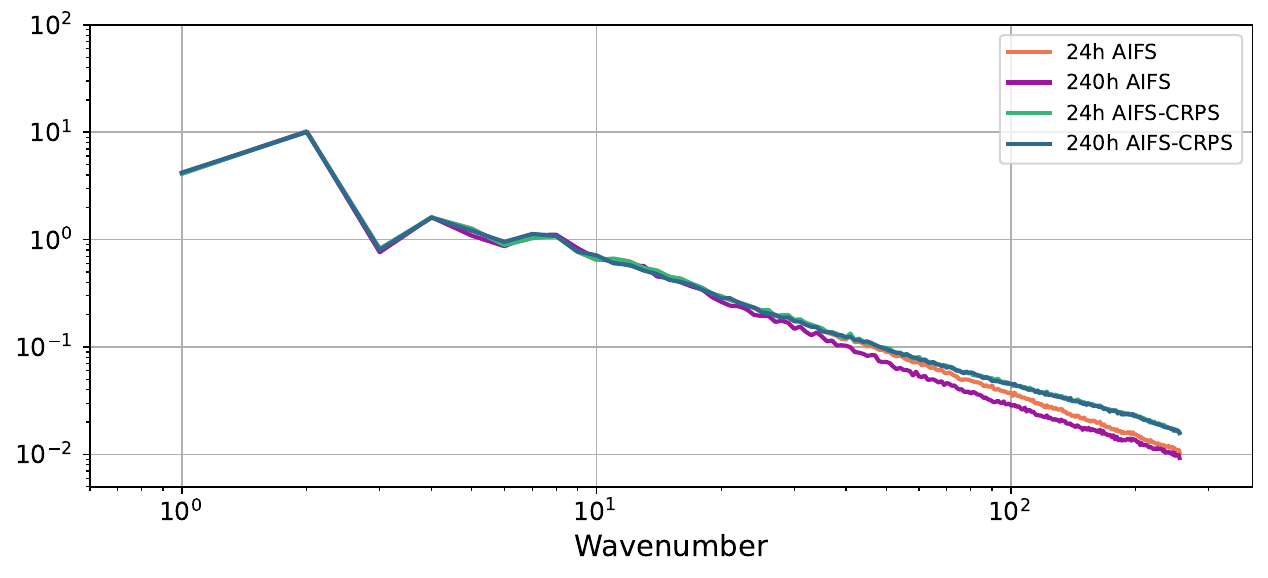}
    \caption{}\label{fig:comp_spc}
\end{subfigure}
\caption{Spectra of geopotential at 500~hPa (\subref{fig:spc_z_1}, \subref{fig:spc_z_2}) and temperature at 850~hPa (\subref{fig:spc_t_2}, \subref{fig:comp_spc}) for different lead times. Step 0~h refer to the initial conditions / IFS analysis. Shown are the AIFS-CRPS ensemble without (\subref{fig:spc_z_1}) and with reference field truncation (\subref{fig:spc_z_2}, \subref{fig:spc_t_2}, \subref{fig:comp_spc}), and AIFS (\subref{fig:comp_spc}). Spectra are averaged over 12 initial dates and the first 8 ensemble members (\subref{fig:spc_z_1}, \subref{fig:spc_z_2} and \subref{fig:spc_t_2}). For the AIFS and AIFS-CRPS comparison (\subref{fig:comp_spc}), the spectra are averaged over 12 initial dates and AIFS-CRPS perturbed member 1 only. For more explanation, please see the text.}
\label{fig:spec}
\end{figure}

The improved representation of variability is also visible in the spectra of 500~hPa geopotential. Small scale variability increases with lead time and propagates to larger scales without reference field truncation (figure~\ref{fig:spc_z_1}). With reference field truncation, there is still a slight increase of small-scale variability relative to the IFS analysis / initial conditions, but the spectra are stable and do not change significantly with lead time (figure~\ref{fig:spc_z_2}). Spectra of less smooth forecast fields are quite stable in general (see figure~\ref{fig:spc_t_2}). For all fields, there is no dampening of smaller scales with lead time visible. This is in contrast to AIFS trained with an MSE loss, where the forecast fields progressively lose energy at higher wavenumbers with lead time (figure~\ref{fig:comp_spc}).

\subsection{Medium-Range}\label{mediumrange}
We compare 50-member 15-day AIFS-CRPS ensemble forecasts with the 9-km 50-member ECMWF IFS (Integrated Forecasting System) ensemble forecasts \citep{lang81380}. Both forecast systems are initialised from the operational IFS ensemble initial conditions and verified against the operational ECMWF analysis. In addition, to the analysis-based verification, forecasts are compared against radiosonde observations of geopotential, temperature and wind speed and SYNOP observations of 2~m temperature, 10~m wind and 24~h total precipitation.

% TODO .. update
\begin{figure}[htpb]
\centering
\begin{subfigure}{0.32\linewidth}
    \includegraphics[width=\linewidth]{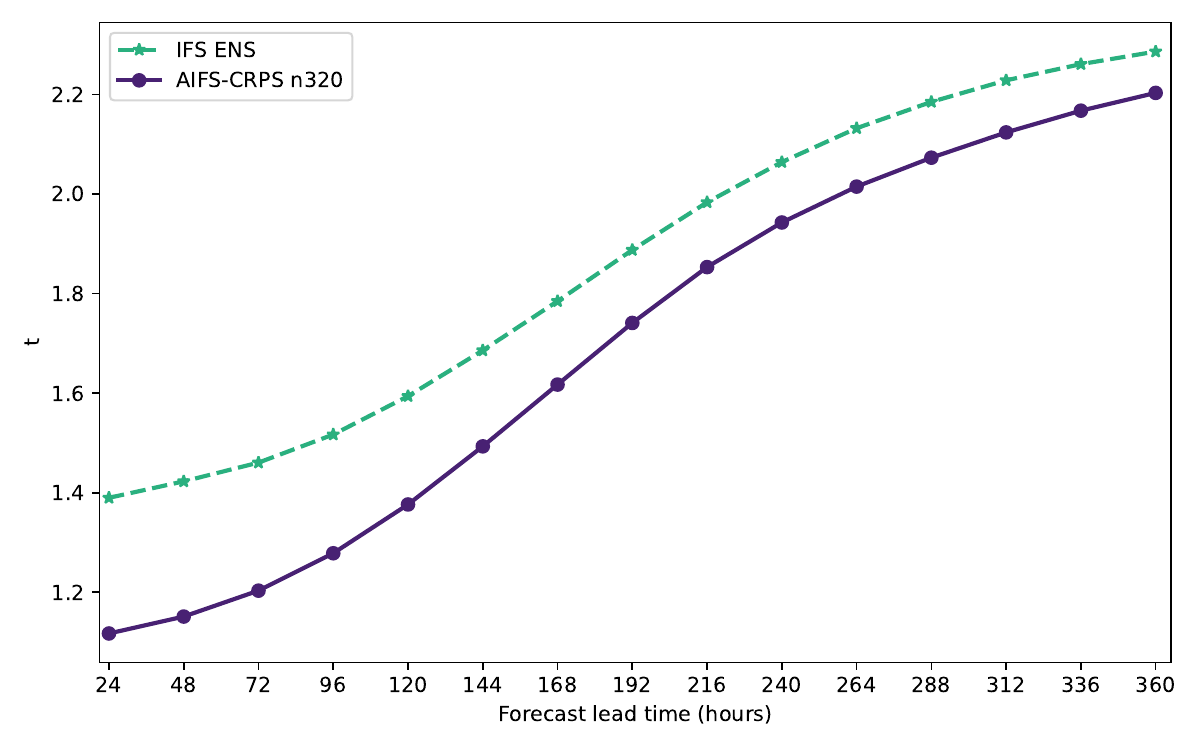}\caption{}\label{fig:crps_2tsfcob_n}
\end{subfigure}
\begin{subfigure}{0.32\linewidth}
    \includegraphics[width=\linewidth]{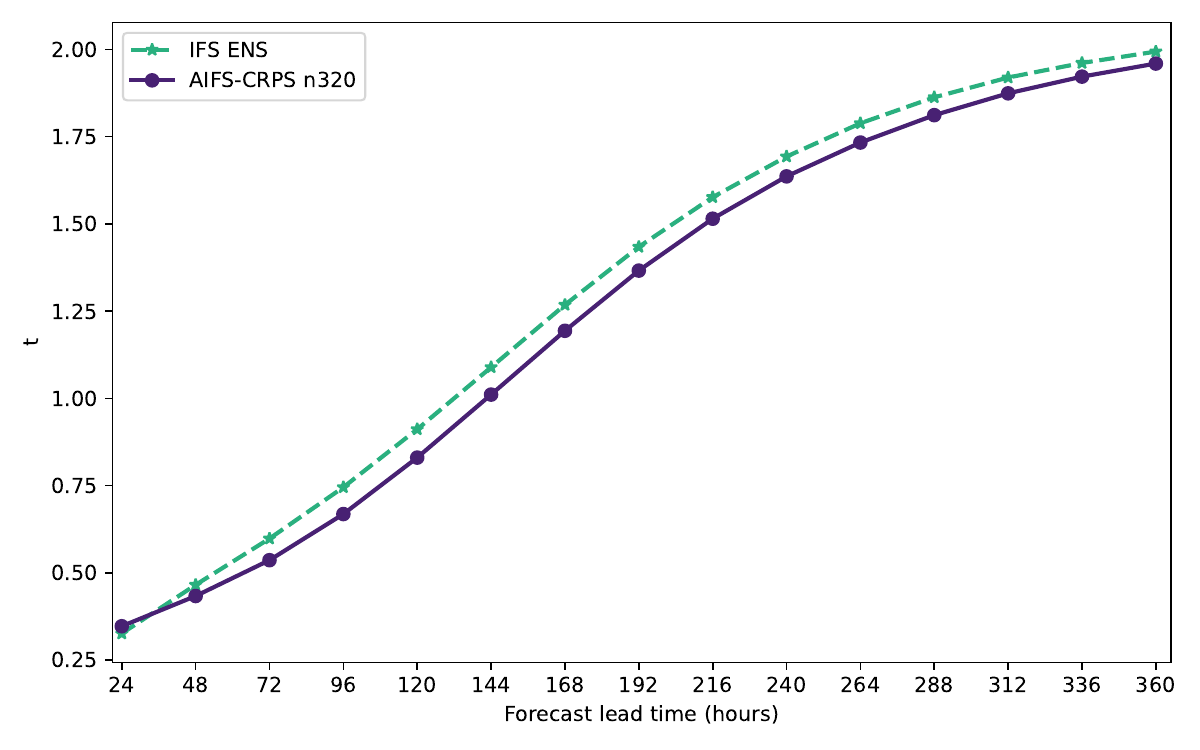}\caption{}\label{fig:crps_t850pl_n}
\end{subfigure}
\begin{subfigure}{0.32\linewidth}
    \includegraphics[width=\linewidth]{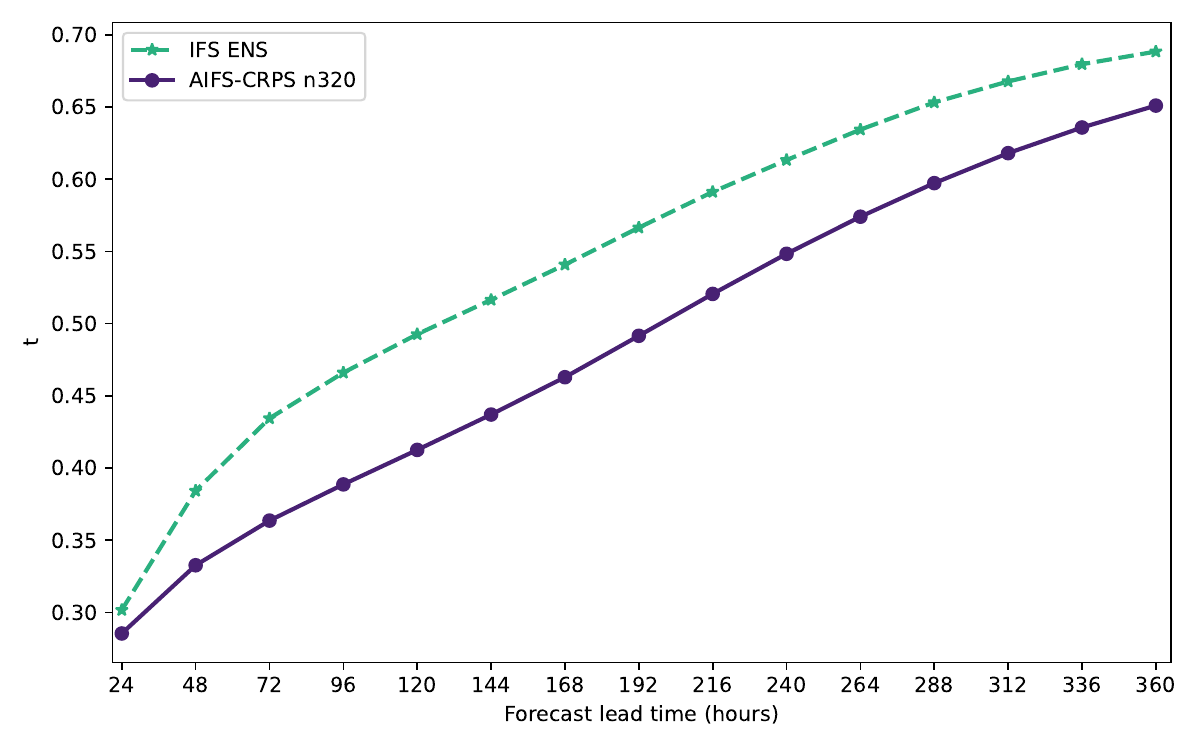}\caption{}\label{fig:crps_t850pl_tropics}
\end{subfigure}
\caption{AIFS-CRPS N320 (blue, solid line) and IFS ensemble (green, dashed line) CRPS of 2~m temperature for different lead times in the nothern extra-tropics verified against SYNOP observations (\subref{fig:crps_2tsfcob_n}), temperature at 850~hPa, northern extra-tropics (\subref{fig:crps_t850pl_n}) and Tropics (\subref{fig:crps_t850pl_tropics}) verified against analyses. Scores are averaged over the period 1 February to 30 September 2024, with forecasts initialised at 00 and 12 UTC.}
\label{fig:crps_example}
\end{figure}
AIFS-CRPS ensemble forecasts  are considerably more skilful than the IFS ensemble for a large number of variables, for example 2m temperature (figure~\ref{fig:crps_2tsfcob_n}), or temperature at 850~hPa (figure \ref{fig:crps_t850pl_n} and \subref{fig:crps_t850pl_tropics}). 

\begin{figure}[htpb]
\centering
\includegraphics[width=0.98\linewidth]{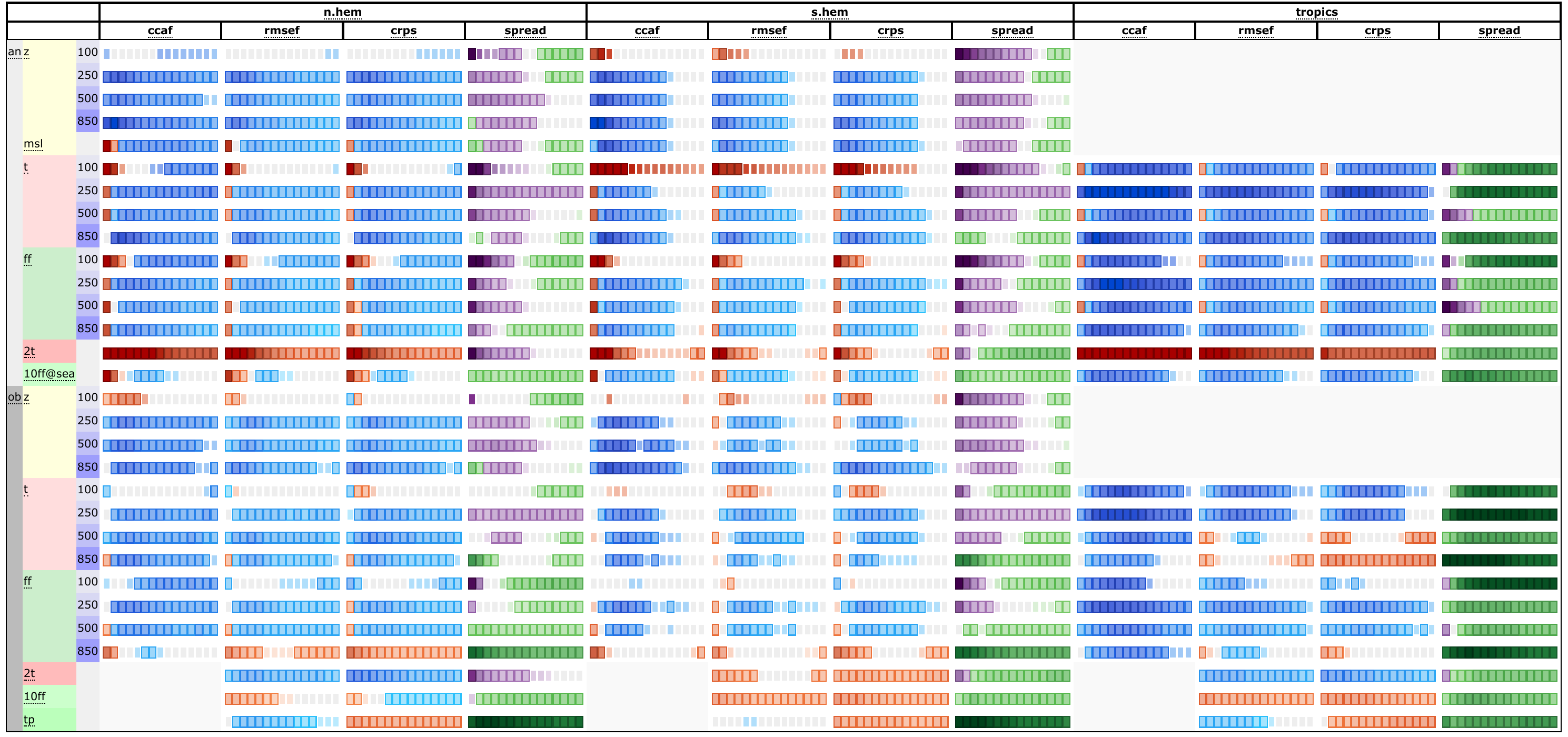}
\caption{Scorecard comparing forecast scores of AIFS-CRPS O96 ensemble (approximately 1.0$\degree$ spatial resolution) versus the IFS ensemble (approximately 0.1$\degree$ spatial resolution), 1 February to 30 September 2024. Forecasts are initialised at 00 and 12 UTC. Shown are relative score changes as function of lead time (day 1 to 15) for northern extra-tropics (n.hem), southern extra-tropics (s.hem) and tropics. Blue colours mark score improvements and red colours score degradations. Purple colours indicate an increase in ensemble standard deviation, while green colours indicate a reduction. Differences that reach 95$\%$ significance level are shown in light shading and differences that reach 99.7$\%$ significance level are shown in dark shading. Variables are geopotential (z), temperature (t), wind speed (ff), mean sea level pressure (msl), 2~m temperature (2t), 10~m wind speed (10ff) and 24 hr total precipitation (tp). Numbers behind variable abbreviations indicate variables on pressure levels (e.g., 500~hPa), and prefix indicates verification against IFS NWP analyses (an) or radiosonde and SYNOP observations (ob). Scores shown are ensemble mean anomaly correlation (ccaf), CRPS, ensemble mean RMSE (rmsef) and ensemble standard deviation (spread).}
\label{fig:scorecard_o96}
\end{figure}

\begin{figure}[htpb]
    \centering
    \includegraphics[width=0.98\linewidth]{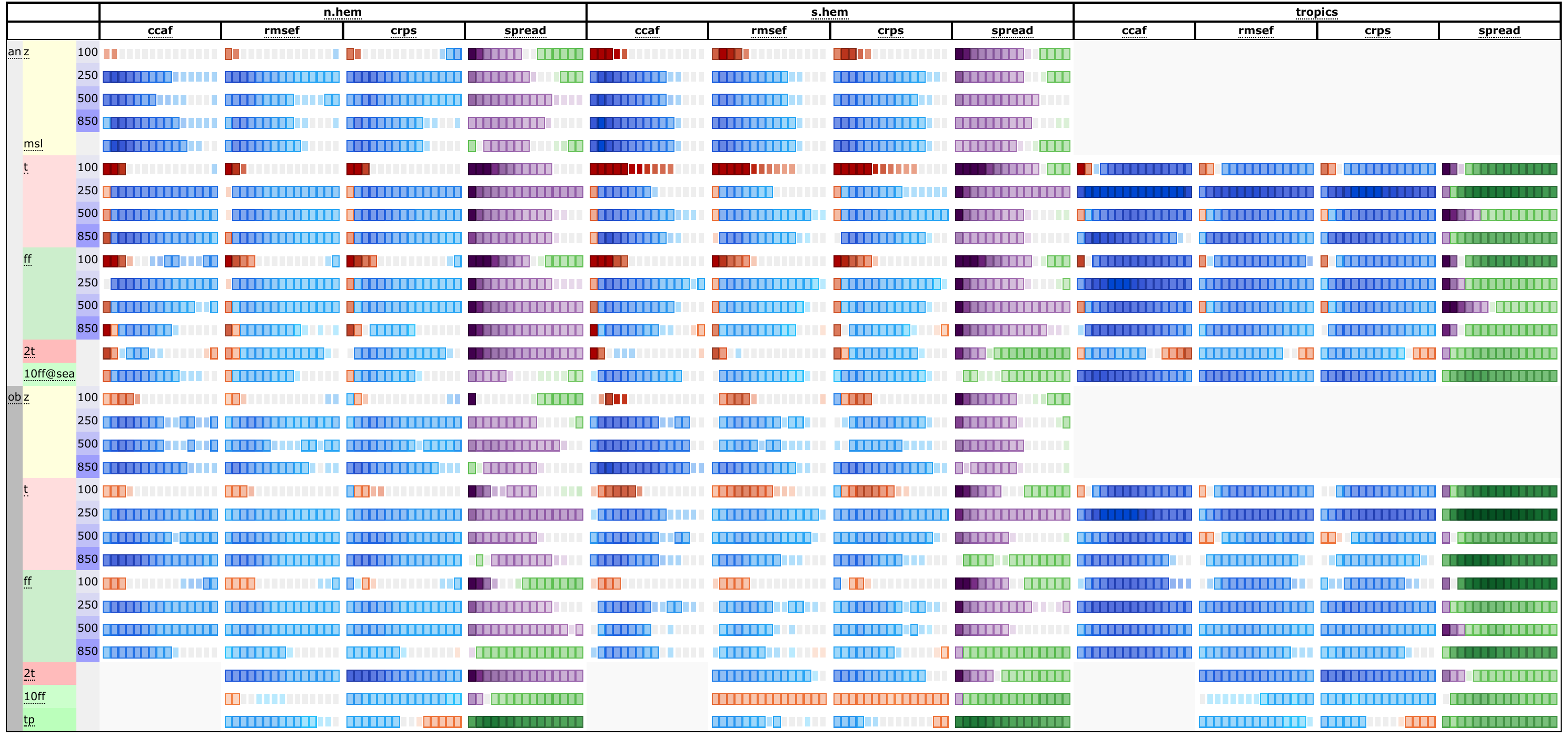}
    \caption{Like Figure \ref{fig:scorecard_o96}, but comparing forecast scores of AIFS-CRPS N320 ensemble (approximately 0.25$\degree$ spatial resolution) versus the IFS ensemble (approximately 0.1$\degree$ spatial resolution). Blue colours mark score improvements and red colours score degradations of AIFS-CRPS N320 compared to the IFS ensemble. Purple colours indicate an increase in ensemble standard deviation, while green colours indicate a reduction.}\label{fig:scorecard_n320}
\end{figure}

Figure~\ref{fig:scorecard_o96} and \ref{fig:scorecard_n320} display scorecards of the relative difference between the AIFS-CRPS and the IFS ensemble forecasts in terms of CRPS, ensemble mean root mean squared error (RMSE), ensemble mean anomaly correlation and ensemble standard deviation (ensemble spread) across a range of variables. Following standard practice, the upper-air variables are interpolated to a 1.5$\degree$ latitude-longitude grid for the verification against analyses.

AIFS-CRPS shows higher forecast skill than the IFS ensemble for most upper air variables, such as 500~hPa geopotential, 250~hPa wind speed. This is reflected by a lower CRPS and RMSE, and higher anomaly correlation. This can be seen for the AIFS-CRPS ensemble with a O96 input grid (figure~\ref{fig:scorecard_o96}) and the AIFS-CRPS ensemble with an N320 grid (figure~\ref{fig:scorecard_n320}). Here the forecast improvements are in the range of 5-20$\%$. Higher up in the atmosphere (100~hPa and above) forecast scores can be degraded compared to the IFS ensemble. This is more pronounced for the AIFS-CRPS N320 ensemble that was trained with a linear pressure scaling only (see section~\ref{sec:experiments}). 

Ensemble spread tends to be larger in the extra-tropics in the first half of the forecast range, in case of the AIFS-CRPS O96 ensemble and for most of the forecast range in the case of the AIFS-CRPS N320 ensemble. In the tropics however, ensemble spread is notably smaller in the AIFS-CRPS than in the IFS ensemble, apart from the first days of the forecast. This is accompanied by a markedly reduced RMSE of the ensemble mean. The ensemble spread of most surface variables is considerably reduced for the AIFS-CRPS N320 ensemble compared to the IFS ensemble, apart from 2m temperature in the northern extra-tropics.

When verified against surface observations, forecasts from the AIFS-CRPS O96 ensemble are more skilful than the IFS ensemble for some variables, e.g. 2m temperature in northern hemisphere and tropics, but they are less skilful for others, like total precipitation. Here, increased resolution plays a role, and the AIFS-CRPS N320 ensemble has higher skill than IFS ensemble for most surface variables. Total precipitation is more skilful than the IFS ensemble for approximately the first 8 days of the forecast and less skilful towards the end of the forecast, likely due to the reduced spread of the AIFS-CRPS ensemble compared to the IFS ensemble.

\begin{figure}[htpb]
    \centering
    \includegraphics[width=0.98\linewidth]{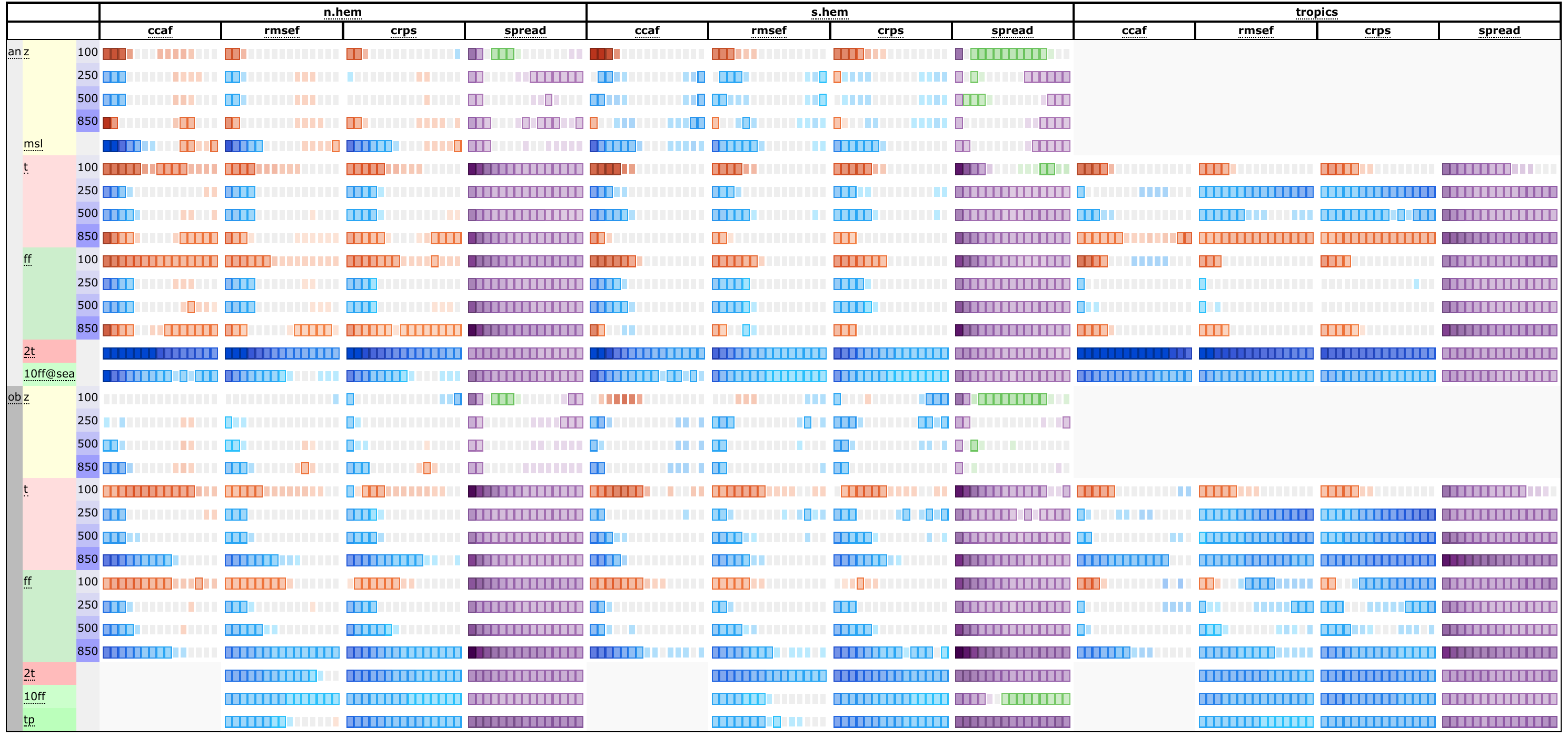}
    \caption{Like Figure \ref{fig:scorecard_o96}, but comparing forecast scores of AIFS-CRPS N320 ensemble (approximately 0.25$\degree$ spatial resolution) versus AIFS-CRPS O96 ensemble (approximately 1.0$\degree$ spatial resolution). Blue colours mark score improvements and red colours score degradations of AIFS-CRPS N320 compared to AIFS-CRPS O96. Purple colours indicate an increase in ensemble standard deviation, while green colours indicate a reduction.}\label{fig:diff_n320_o96}
\end{figure}

The scorecard shown in figure~\ref{fig:diff_n320_o96} compares the AIFS-CRPS N320 ensemble with the AIFS-CRPS O96 ensemble. The AIFS-CRPS N320 has higher forecast skill for most variables, especially for surface variables, where differences are large. The ensemble spread is also increased for most variables, especially surface variables. There is some degradation at 100~hPa and above, related to the differences in pressure scaling applied in the loss function. Wind speed and temperature at 850~hPa appear degraded in the AIFS-CRPS N320 ensemble compared to the AIFS-CRPS O96 ensemble, when verified against analyses. However, when verified against observations, the AIFS-CRPS N320 ensemble shows large improvements for these variables.

\begin{figure}[htpb]
    \centering
    \begin{subfigure}{0.32\linewidth}
        \includegraphics[width=\linewidth]{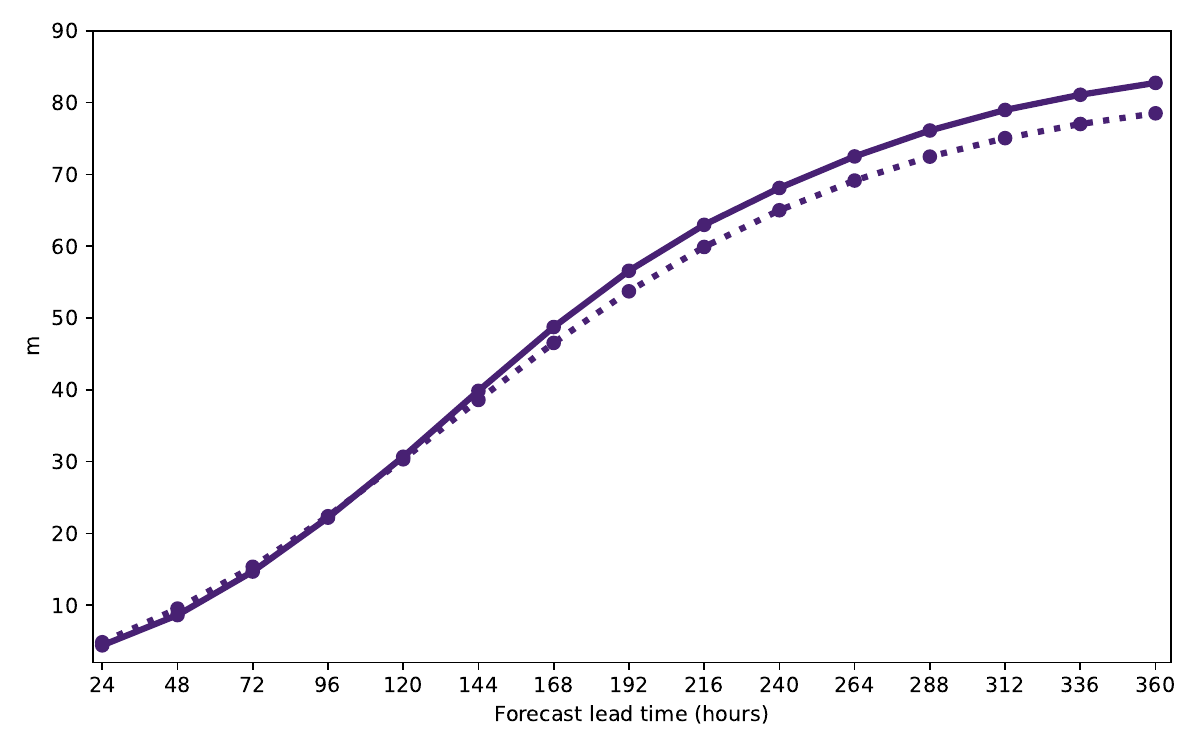}\caption{}\label{fig:ifszsprerr}
    \end{subfigure}
    \begin{subfigure}{0.32\linewidth}
        \includegraphics[width=\linewidth]{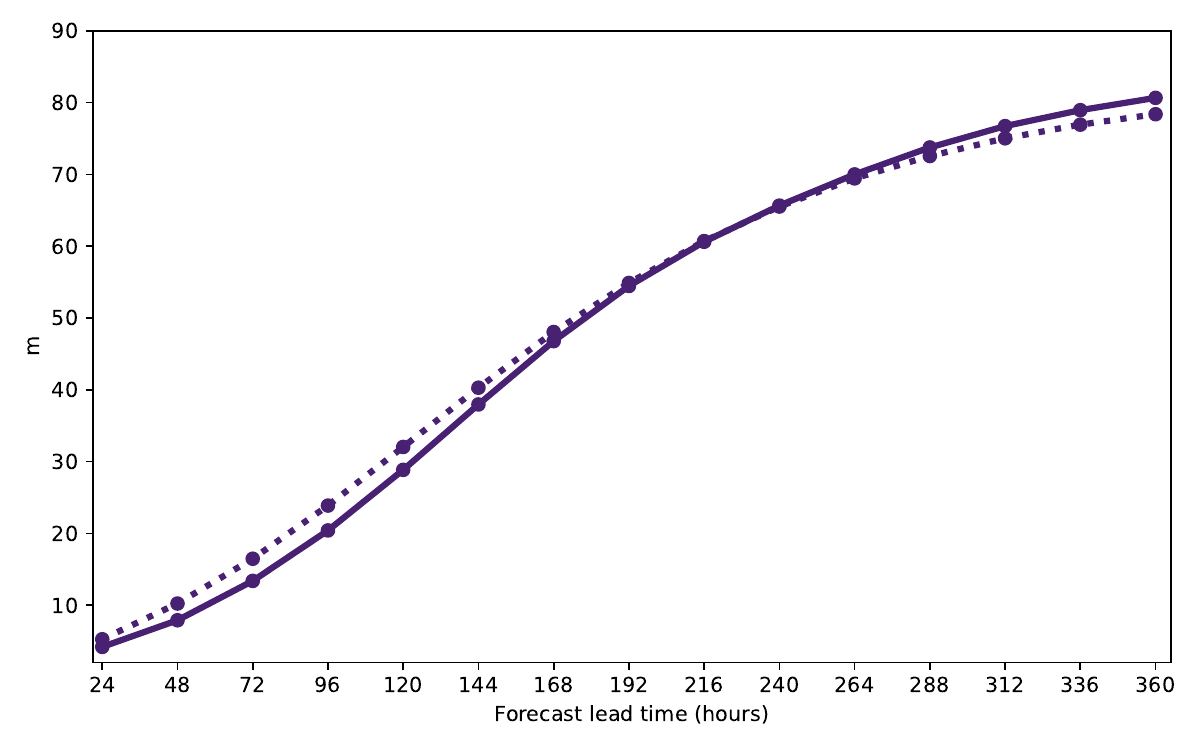}\caption{}\label{fig:o96zsprerr}
    \end{subfigure}
    \begin{subfigure}{0.32\linewidth}
        \includegraphics[width=\linewidth]{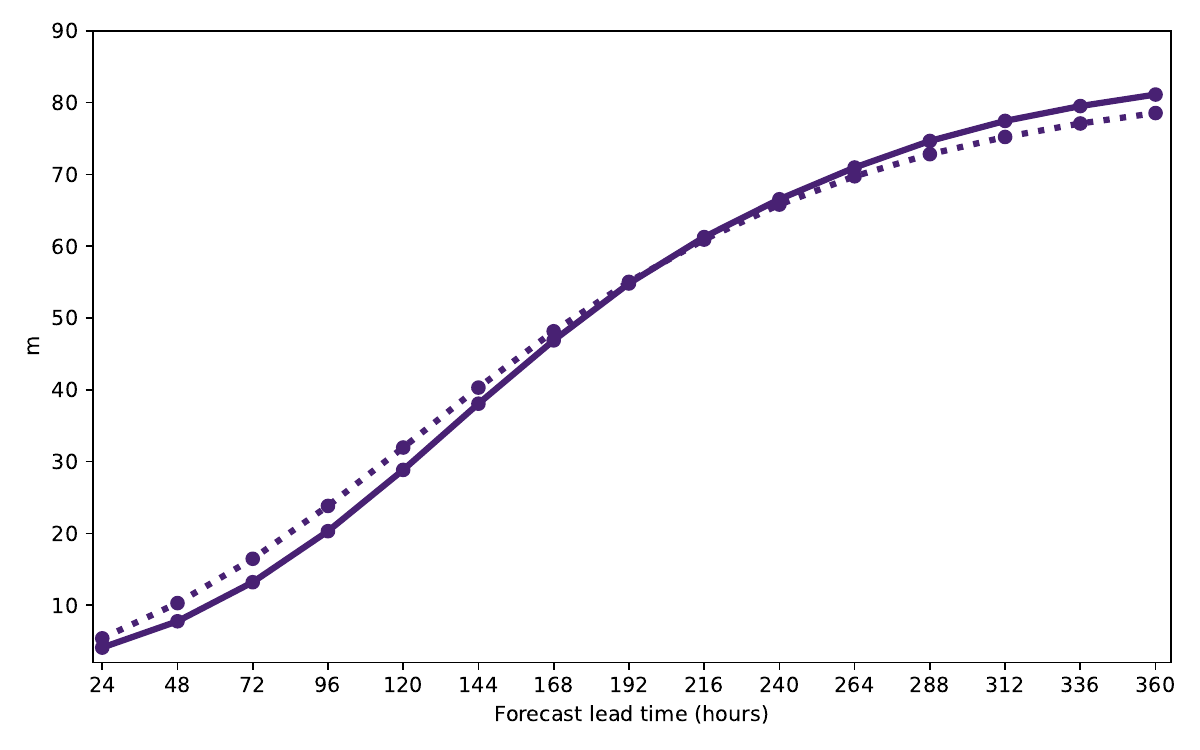}\caption{}\label{fig:n320zsprerr}
    \end{subfigure}
    \begin{subfigure}{0.32\linewidth}
        \includegraphics[width=\linewidth]{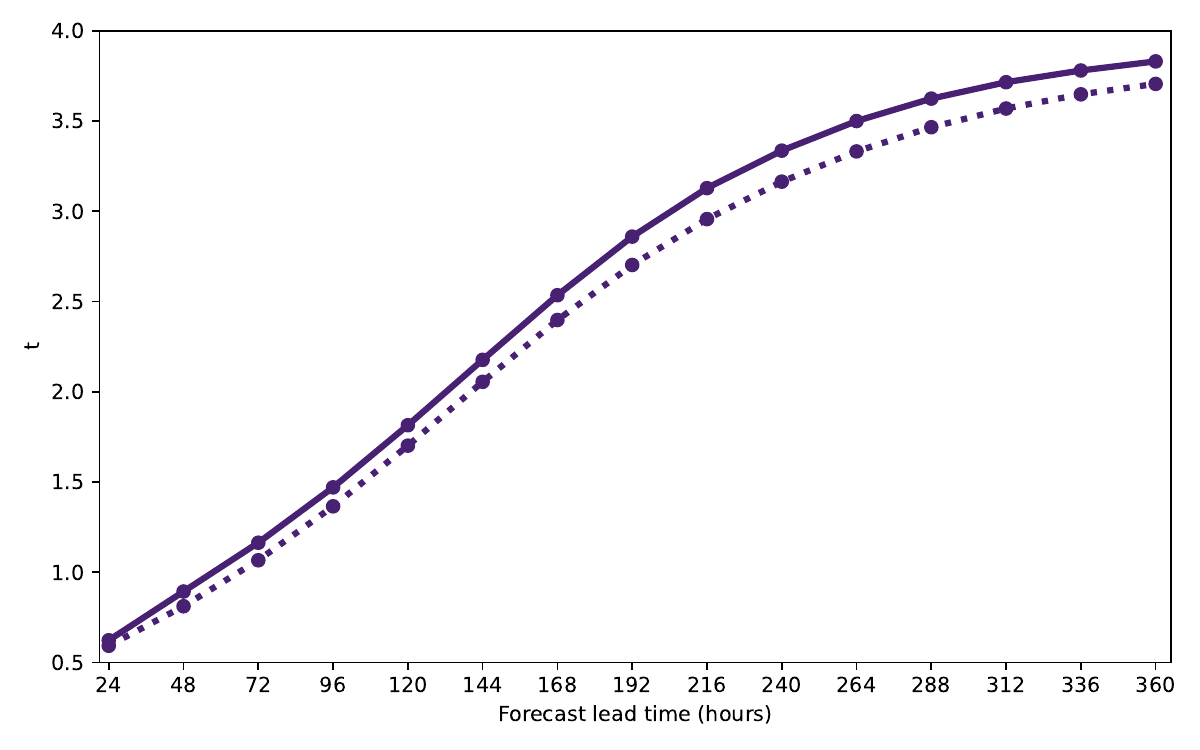}\caption{}\label{fig:ifstsprerr}
    \end{subfigure}
    \begin{subfigure}{0.32\linewidth}
        \includegraphics[width=\linewidth]{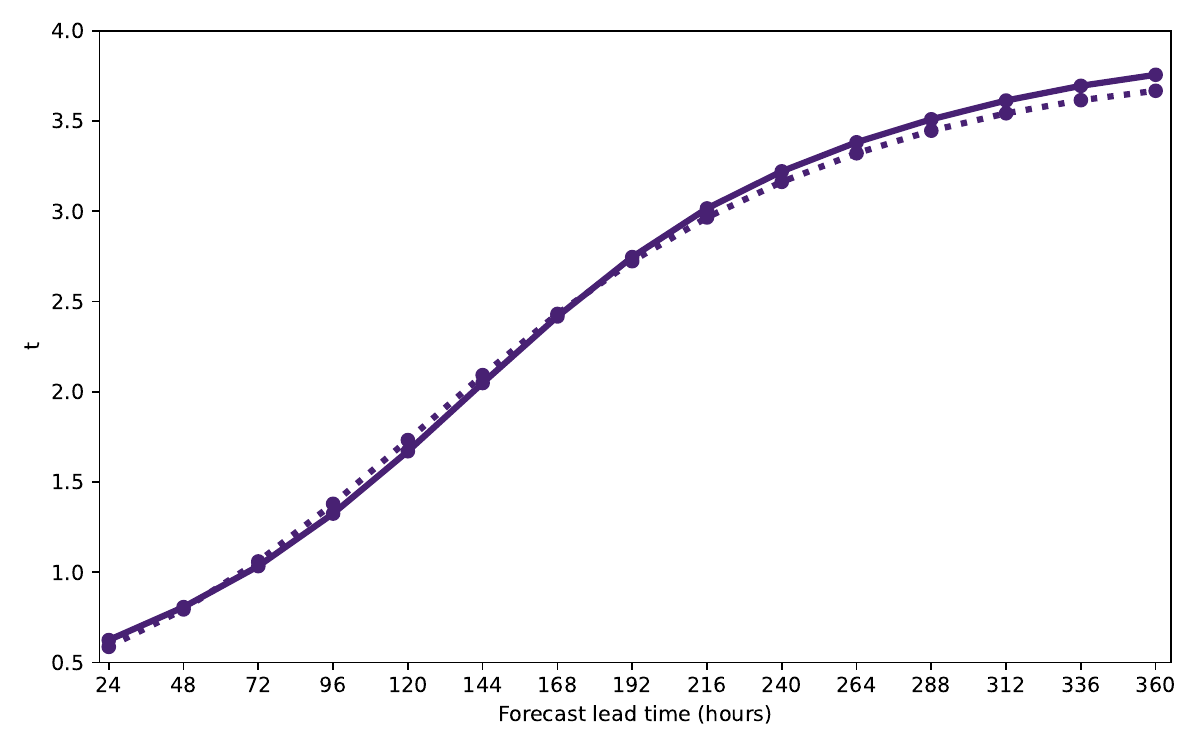}\caption{}\label{fig:o96tsprerr}
    \end{subfigure}
    \begin{subfigure}{0.32\linewidth}
        \includegraphics[width=\linewidth]{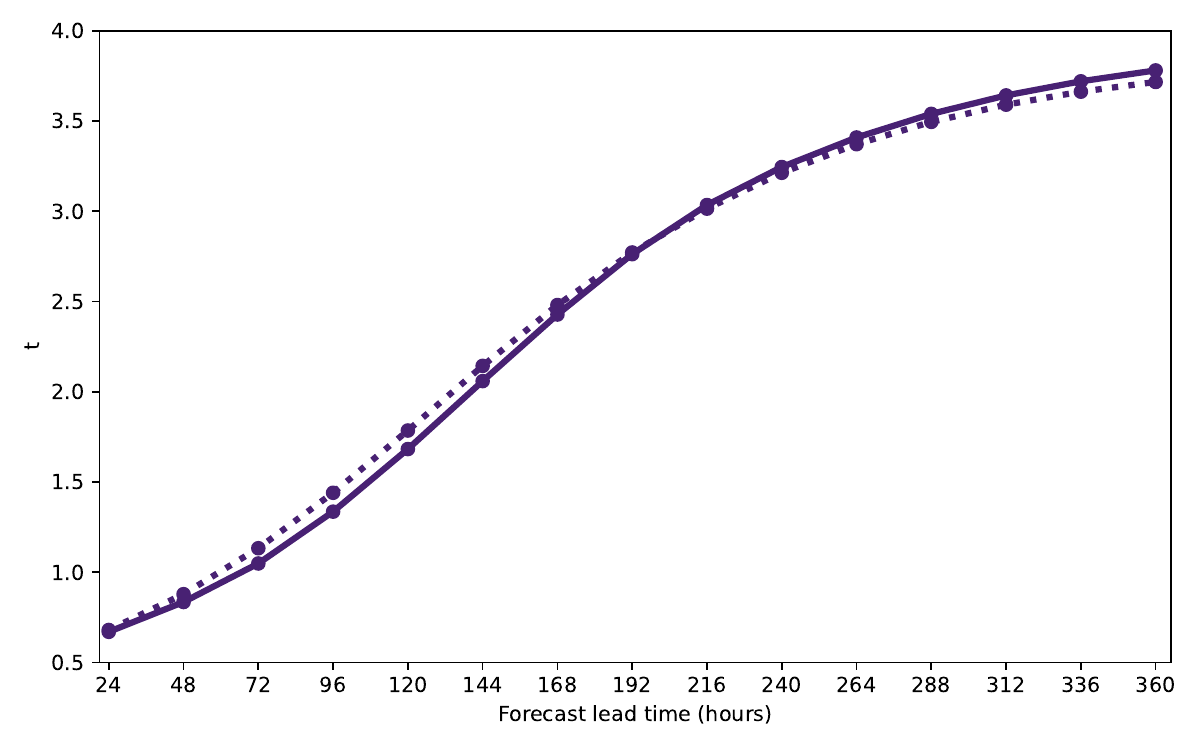}\caption{}\label{fig:n320tsprerr}
    \end{subfigure}
    \caption{Ensemble mean RMSE (solid line) and ensemble spread (dotted line) for geopotential at 500~hPa (\subref{fig:ifszsprerr}, \subref{fig:o96zsprerr} and \subref{fig:n320zsprerr}) and temperature at 850~hPa (\subref{fig:ifstsprerr}, \subref{fig:o96tsprerr}, \subref{fig:n320tsprerr}) in the northern extra-tropics. Shown are IFS ensemble (\subref{fig:ifszsprerr}, \subref{fig:ifstsprerr}), AIFS-CRPS O96 (\subref{fig:o96zsprerr}, \subref{fig:o96tsprerr}) and N320 (\subref{fig:n320zsprerr}, \subref{fig:n320tsprerr}).}
    \label{fig:spred-error}
\end{figure}

When comparing ensemble mean RMSE and ensemble spread, it is apparent that AIFS-CRPS tends to be over-dispersive in the extra-tropics for a range of variables. The ensemble spread is larger than the ensemble mean RMSE. The over-dispersion is especially visible for geopotential at 500~hPa. The correspondence between ensemble spread and ensemble mean RMSE is worse than for the IFS ensemble (compare figure~\ref{fig:ifszsprerr}, \subref{fig:o96zsprerr} and \subref{fig:n320zsprerr}). For temperature at 850~hPa, the ensemble spread of the AIFS-CRPS O96 ensemble is between the IFS ensemble and the AIFS-CRPS ensemble (compare figure~\ref{fig:ifstsprerr}, \subref{fig:o96tsprerr} and \subref{fig:n320tsprerr}).

\subsection{Subseasonal}\label{section:s2s}
Although AIFS-CRPS is designed and trained primarily for medium-range ensemble forecasting, it is also stable at longer lead times and competitive with state-of-the-art subseasonal forecasts. Subseasonal-to-seasonal (S2S) forecasts provide an overview of potential global and regional weather patterns at lead times of two weeks to two months and fill the gap between medium-range weather forecasts and long-range seasonal outlooks \citep{vitart2008vareps, white2017s2s, vitart2018s2s}. The predictability at S2S timescales is largely determined by atmospheric initial conditions, though there are also important contributions from slowly evolving components of the Earth System, including the oceans, sea-ice, and land-surface properties \citep{meehl2021initialized}.
 
The AIFS-CRPS subseasonal reforecast dataset is comprised of 46-day, 8-member ensemble forecasts initialized once per week over the period 2018-2022 for a total of 260 start dates. Here, we assess the performance of the AIFS-CRPS O96 ensemble. Initial conditions are from ERA5 data with perturbations derived from the ERA5 ensemble of data assimilations (EDA). To provide context to the subseasonal performance of AIFS-CRPS, we compare against operational IFS reforecasts produced during 2023, which we subset to use the same ensemble size and start dates as AIFS-CRPS. The operational IFS reforecasts were also initialized from ERA5 but with perturbations derived using a combination of the ERA5 EDA and singular vectors. Further information on the performance and configuration of IFS subseasonal reforecasts is available in \citet{roberts2023euro}. 

\begin{figure}[htpb]
\centering
\begin{subfigure}{0.45\linewidth}
    \includegraphics[width=\linewidth]{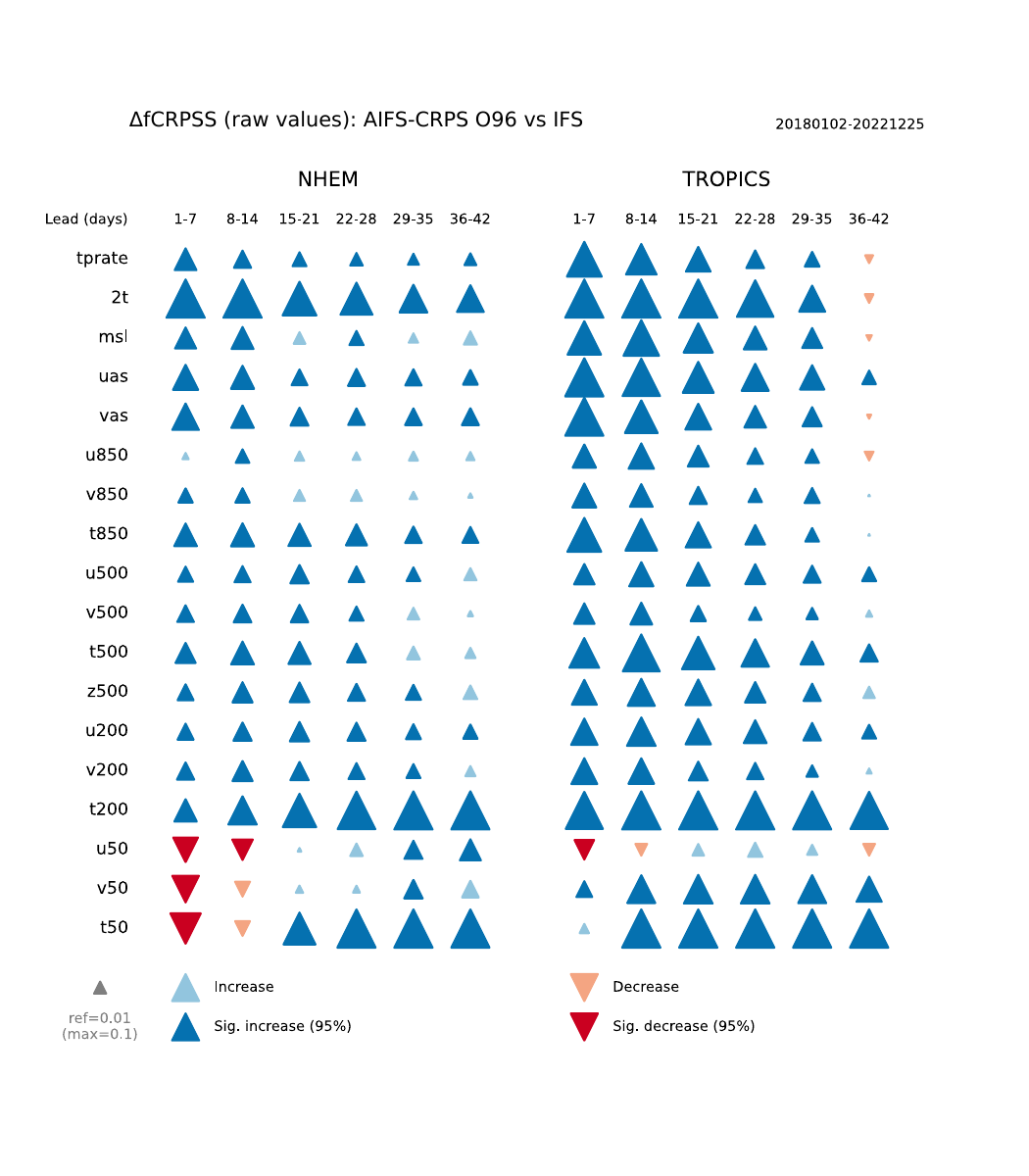}\caption{}
    \label{fig:weekly_raw}
\end{subfigure}
\hfill
\begin{subfigure}{0.45\linewidth}
    \includegraphics[width=\linewidth]
    {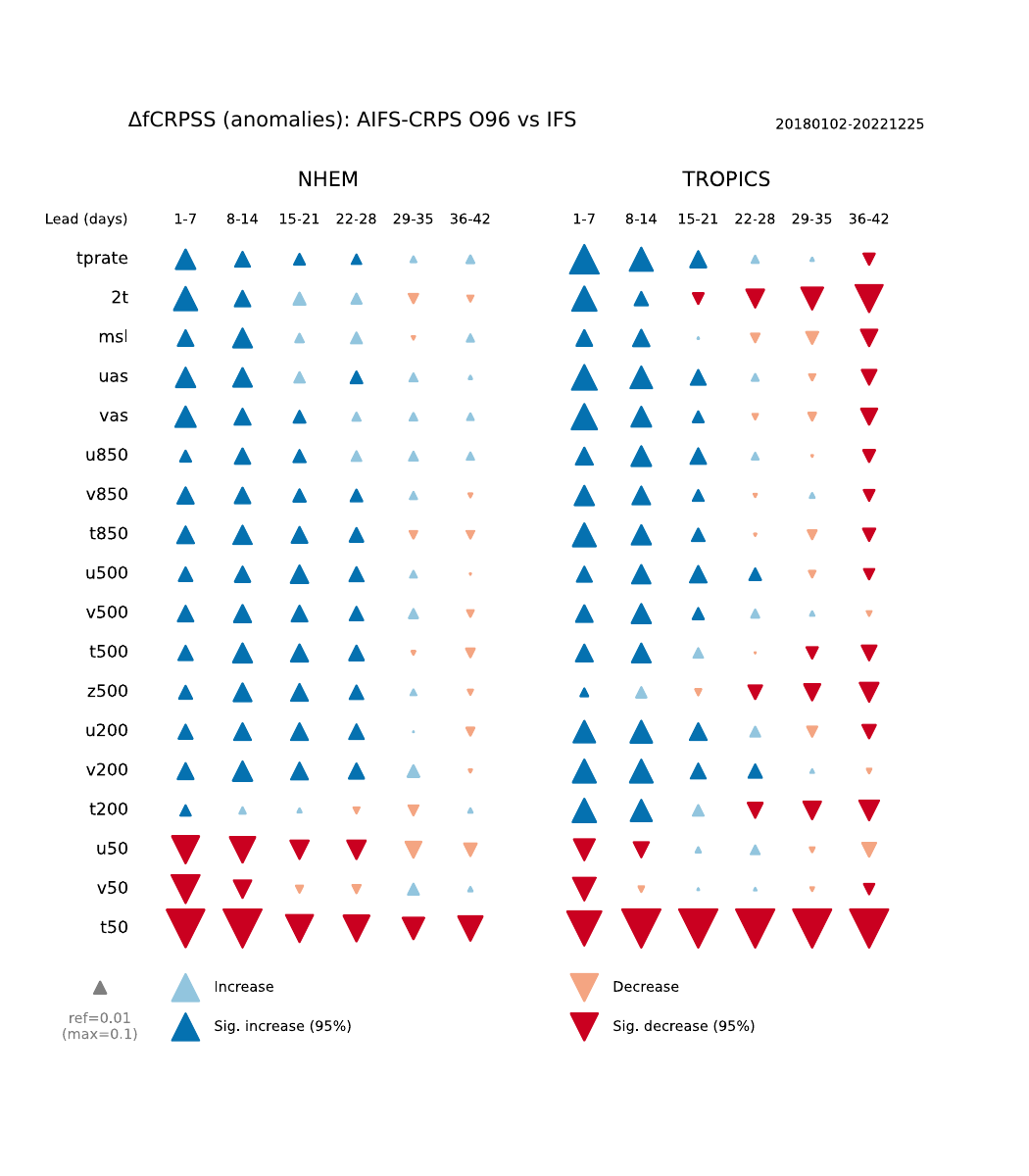}
\caption{}
    \label{fig:weekly_ano}
\end{subfigure}
\caption{Score cards summarizing differences between AIFS-CRPS and IFS in the fair continuous ranked probability skill score (fCRPSS) for raw weekly mean data (\subref{fig:weekly_raw}) and weekly mean anomalies (\subref{fig:weekly_ano}) defined relative to start-date and lead-time climatologies (see \cite{roberts2024unbiased}). Scores are aggregated over the northern hemisphere (30$^{\circ}$N-90$^{\circ}$N) and tropics (30$^{\circ}$S-30$^{\circ}$N) on a regular 2.5$^{\circ}$$\times$2.5$^{\circ}$ latitude-longitude grid. Differences in fCRPSS are defined as $\Delta \textnormal{fCRPSS} = \frac{ \textnormal{fCRPS}_{IFS} -  \textnormal{fCRPS}_{AIFS}} {\textnormal{CRPS}_{Clim}}$, where $\textnormal{fCRPS}_{IFS}$ and $\textnormal{fCRPS}_{AIFS}$ are the weighted-mean fair CRPS of IFS and AIFS-CRPS reforecasts, respectively, and $\textnormal{CRPS}_{Clim}$ is the weighted-mean CRPS of reference forecasts constructed from the climatological distribution of observed values. Positive (blue) triangles indicate that $\Delta \textnormal{fCRPSS}$ is increased and thus AIFS-CRPS is improved compared to IFS. Negative (red) triangles indicate that $\Delta \textnormal{fCRPSS}$ is reduced and thus AIFS-CRPS is degraded relative to IFS. Symbol areas are proportional to the magnitude of $\Delta \textnormal{fCRPSS}$ and significance is determined by block bootstrap resampling with start dates pooled by calendar month as described in \citet{roberts2023euro}. The area of the grey reference triangle corresponds to $\Delta \textnormal{fCRPSS} = 0.01$. The variables shown are 2m temperature (2t), total precipitation rate (tprate), mean sea level pressure (msl), 10m zonal and meridional wind (uas/vas), temperature (t), zonal/meridional wind (u/v), and geopotential height (z). Numbers in variable names correspond to pressure levels in hPa. Forecasts are verified against ERA5 and all weekly means are constructed to ensure consistent sampling of the available data.}
\label{fig:weekly_fcrpss}
\end{figure}

The subseasonal forecast skill of AIFS-CRPS relative to IFS is summarized in figure \ref{fig:weekly_fcrpss}, which shows differences in the fair continuous ranked probability skill score ($\Delta$fCRPSS) aggregated over different regions, where $\Delta$fCRPSS is defined such that positive values are indicative of higher skill in AIFS-CRPS relative to IFS. To disentangle the impact of changes in the mean state from changes in the predictability of forecast anomalies we calculate changes in weekly mean forecast skill in two different ways. Firstly, we calculate $\Delta$fCRPSS from raw weekly means without any post-processing, such that ($\Delta$fCRPSS) includes the influence of differences in systematic model biases (figure \ref{fig:weekly_raw}). From this comparison, it is evident that forecast skill is improved in AIFS-CRPS compared to IFS for a range of surface and tropospheric parameters at subseasonal lead times. These improvements are also reflected in other scores, such as the RMSE of the ensemble mean (not shown). These differences are particularly evident in the tropics, where they primarily reflect small improvements to the mean state. For example, the mean RMSE of tropical 200 hPa temperatures is $\sim$0.1 K lower in AIFS-CRPS than IFS. 

We also evaluate $\Delta$fCRPSS calculated from weekly mean anomalies (figure \ref{fig:weekly_ano}), where anomalies are defined relative to start-date and lead-time dependent climatologies to minimize the influence of systematic model biases. Crucially, this evaluation is more representative of the potential impact on real-time subseasonal forecasts, which are typically presented as anomalies or tercile probabilities that are defined with respect to climatologies constructed from an associated set of historical reforecasts. To ensure our evaluation is unbiased despite the short reforecast period, we construct reference climatologies separately for each forecast member following `method D' of \citet{roberts2024unbiased}. We also increase the sample size of reference climatologies by using reforecast dates in all other years within $\pm$7 days of the calendar date of the anomaly forecast. For example, forecast anomalies for January 9th 2022 are defined relative to the climatology constructed from all reforecasts initialized on January 2nd, January 9th, January 16th over the period 2018-2021.

It is clear from comparing the `raw' and `anomaly-based' estimates of $\Delta$fCRPSS in figure \ref{fig:weekly_fcrpss} that a large fraction of the differences in subseasonal forecast performance between AIFS-CRPS and IFS originate from differences in the representation of the mean state. Nevertheless, AIFS-CRPS still offers considerable improvements in forecast skill compared to IFS for many surface and tropospheric parameters for lead times of 2-3 weeks. At longer lead times, despite improvements to the mean state, anomaly-based estimates of forecast skill are very similar in AIFS-CRPS and IFS. We see some degradation in week 6 in the Tropics. In addition, anomaly-based forecast skill in the stratosphere is significantly worse in AIFS-CRPS than IFS, despite the minimum pressure scaling (see section~\ref{sec:loss}). This is consistent with the medium-range evaluation in section~\ref{mediumrange} and the reduced weight given to stratospheric fields in the loss calculation. 

To complement our evaluation of weekly mean forecast skill at model grid points (figure~\ref{fig:weekly_fcrpss}), we also evaluate the ability of AIFS-CRPS to make accurate forecasts of the Madden-Julian Oscillation (MJO), which is the leading mode of intraseasonal variability in the tropics \citep{madden1971detection}. To evaluate the predictability of the MJO, we compute an approximation of the Wheeler and Hendon (2004) Real-time Multivariate MJO (RMM) index that is derived from zonal wind anomalies without contributions from outgoing longwave radiation flux anomalies, which are not available from AIFS-CRPS. Other than setting outgoing longwave radiation (OLR) anomalies to zero, the calculation of our surrogate RMM index follows \citet{wheeler2004all} and \citet{gottschalck2010framework} and is the same for all data sources.

\begin{figure}[htpb]
\centering
\begin{subfigure}{0.45\linewidth}
    \includegraphics[width=\linewidth]{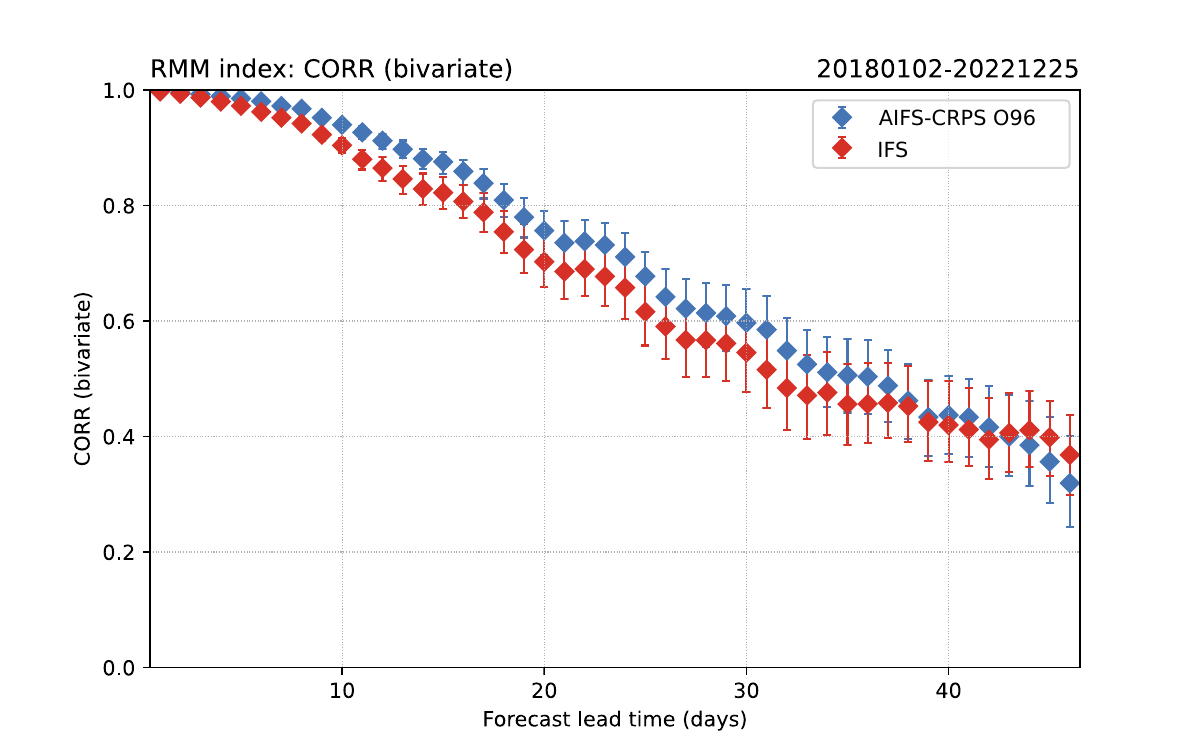}\caption{}
    \label{fig:mjo_bivar_corr}
\end{subfigure}
\hfill
\begin{subfigure}{0.45\linewidth}
    \includegraphics[width=\linewidth]{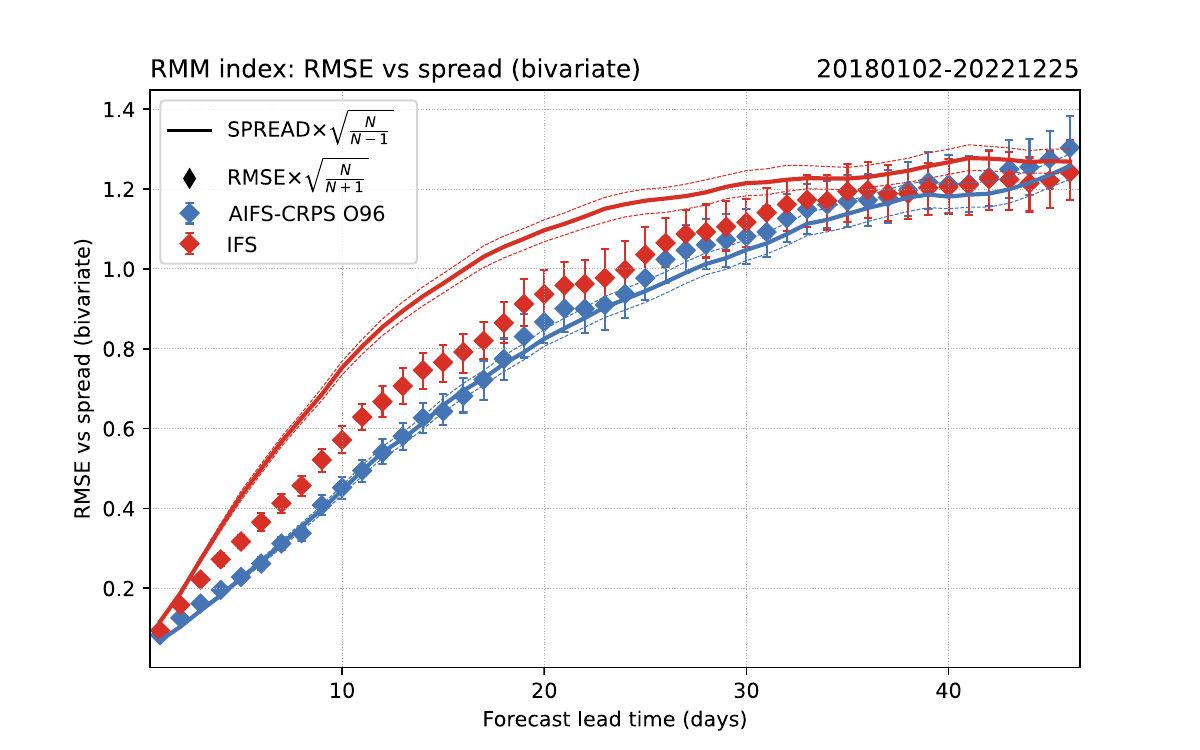}\caption{}
    \label{fig:mjo_bivar_rmse_spread}
\end{subfigure}
\caption{(\subref{fig:mjo_bivar_corr}) Bivariate correlations for an MJO index calculated from 200 hPa and 850 hPa zonal wind anomalies for AIFS-CRPS (blue) and operational IFS reforecasts run in 2023 (red). The MJO index used here is an approximation for the full \citet{wheeler2004all} Real-time Multivariate MJO index as it excludes contributions from outgoing longwave radiation that are not available from AIFS-CRPS. For both systems, correlations are calculated with respect to the same indices calculated from ERA5. Error bars represent the 2.5th and 97.5th percentiles of the distribution created by block-bootstrap resampling of the available start dates. (\subref{fig:mjo_bivar_rmse_spread}) Estimates of root mean square error (RMSE; diamonds) and average ensemble spread (solid lines) for the MJO index described in the text. Spread and RMSE are scaled by factors of $sqrt{\frac{N}{N-1}}$ and $sqrt{\frac{N}{N+1}}$, respectively, to ensure estimates are unbiased with sample size ($N$) as described in \citet{leutbecher2008ensemble}.}
\label{fig:mjo_spread_error_corr}
\end{figure}

Estimates of the MJO skill in AIFS-CRPS and IFS reforecasts are shown in figure \ref{fig:mjo_spread_error_corr}. Despite the short reforecast period, MJO skill from AIFS-CRPS is consistently higher than IFS for several metrics, including correlations, RMSE of the ensemble mean (figure~\ref{fig:mjo_spread_error_corr}), and the fair CRPS (not shown). In addition, AIFS-CRPS exhibits a remarkably good agreement for MJO indices between the average ensemble spread and RMSE of the ensemble mean all lead times, which is required for reliable MJO forecasts (figure~\ref{fig:mjo_bivar_rmse_spread}). The higher correlations and lower RMSE for MJO indices in AIFS-CRPS are not a consequence of an unrealistic representation of MJO amplitude of activity, which might favour deterministic measures of ensemble mean forecast skill. Instead, they seem to be a consequence of genuine improvements to the propagation of MJO-related zonal wind anomalies in the tropics.

\begin{figure}[htpb]
\centering
\begin{subfigure}{0.33\linewidth}
    \includegraphics[width=\linewidth, trim={0cm 4cm 0cm 3cm}, clip]{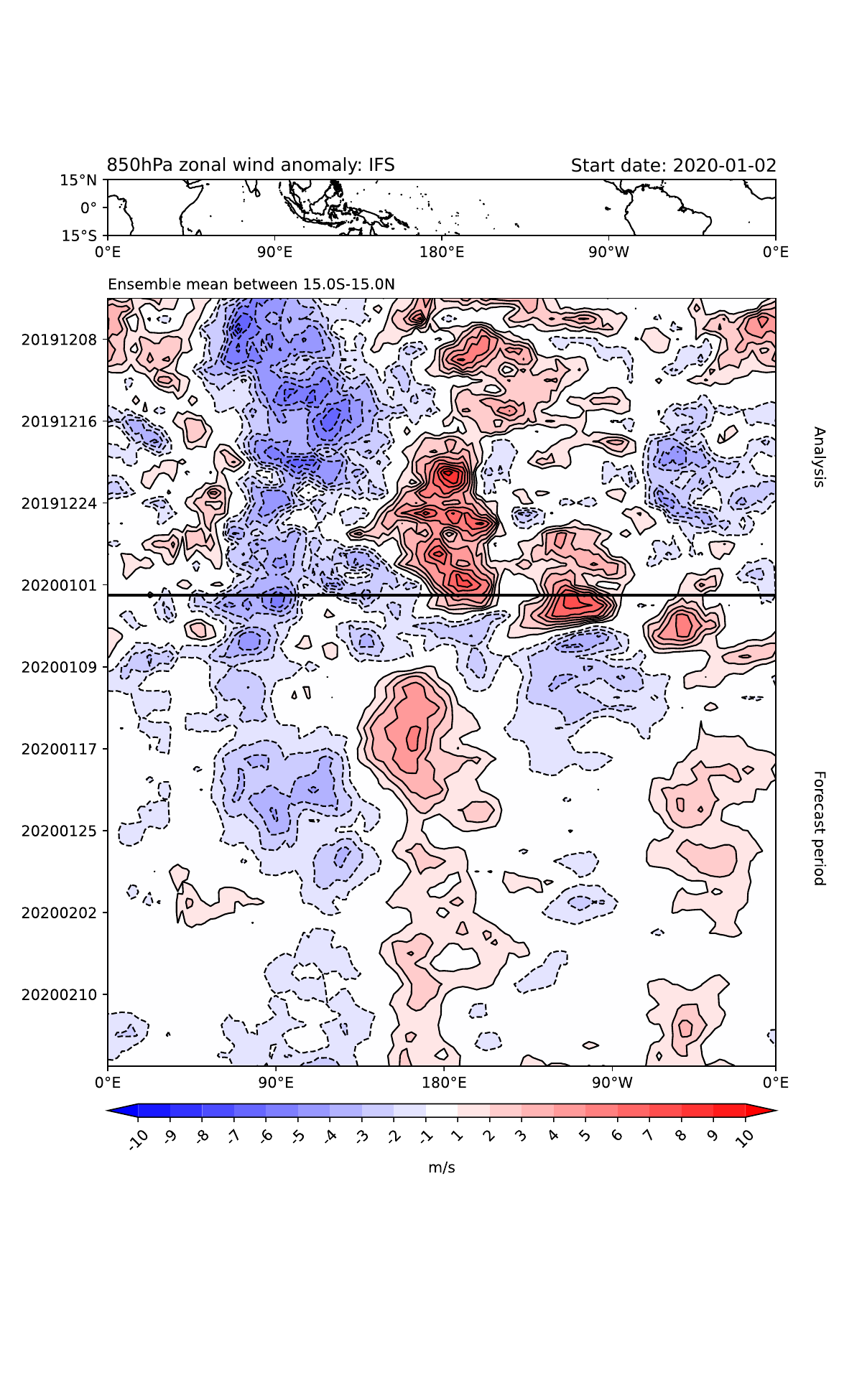}\caption{}
    \label{fig:mjo_case_ifs}
\end{subfigure}
\begin{subfigure}{0.33\linewidth}
    \includegraphics[width=\linewidth, trim={0cm 4cm 0cm 3cm}, clip]{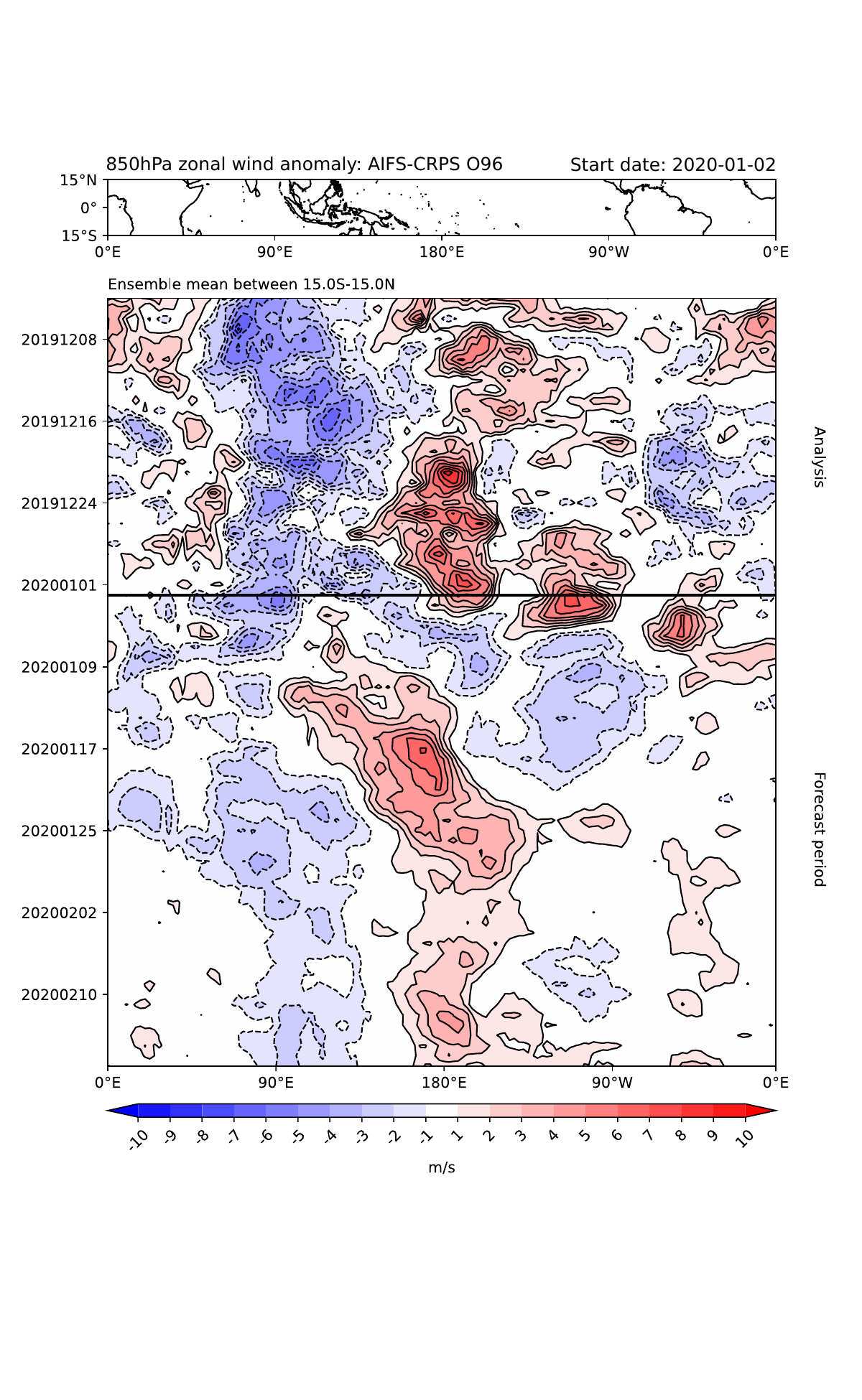}\caption{}
    \label{fig:mjo_case_aifs}
\end{subfigure}
\begin{subfigure}{0.33\linewidth}
    \includegraphics[width=\linewidth, trim={0cm 4cm 0cm 3cm}, clip]{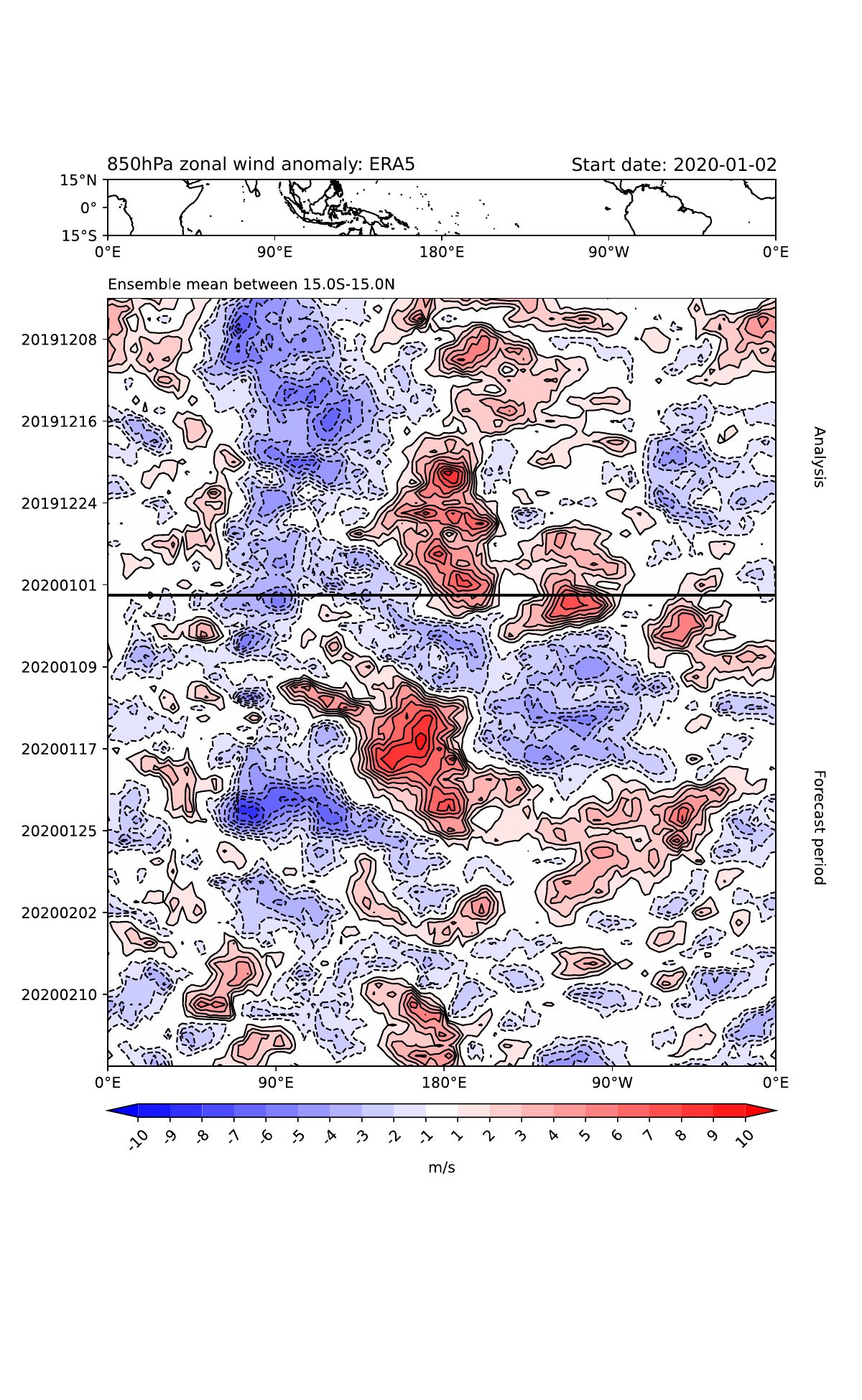}\caption{}
    \label{fig:mjo_case_era5}
\end{subfigure}
\caption{Hovm\"{o}ller diagrams showing the evolution of zonal wind anomalies at 850 hPa meridionally averaged from 15$^{\circ}$S-15$^{\circ}$N. All panels show the evolution of zonal wind anomalies in ERA5 for the 30 days prior to the forecast start date (i.e. data above the grey line). Anomalies below the grey line are from (\subref{fig:mjo_case_ifs}) IFS ensemble mean forecast initialized on 2020-01-02, (\subref{fig:mjo_case_aifs}) AIFS-CRPS ensemble mean forecast initialized on 2020-01-02, and (\subref{fig:mjo_case_era5}) ERA5.}
\label{fig:mjo_hov}
\end{figure}

\begin{figure}[htpb]
\centering
\begin{subfigure}{0.45\linewidth}
    \includegraphics[width=\linewidth]{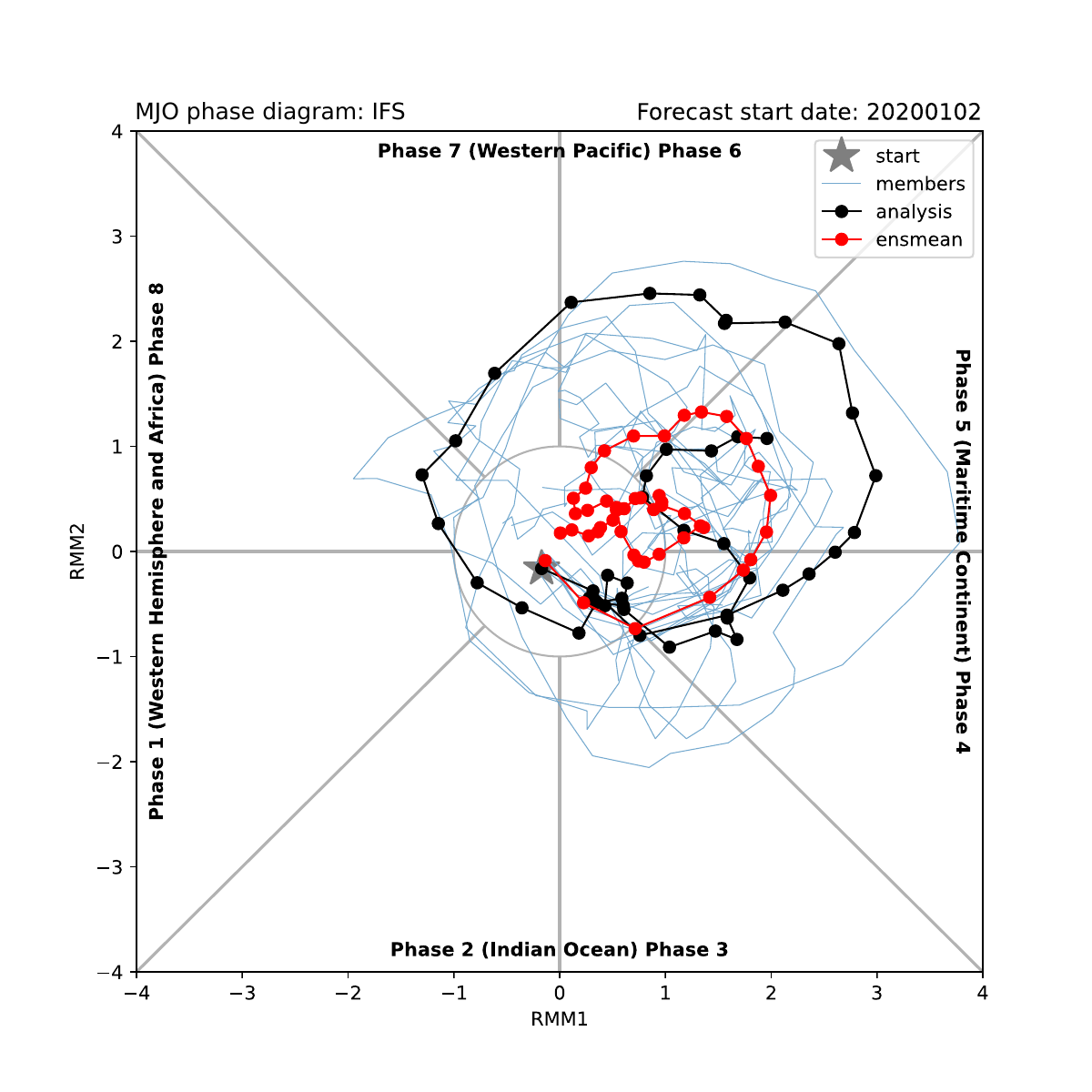}\caption{}
    \label{fig:mjo_case_ifs_phase}
\end{subfigure}
\hfill
\begin{subfigure}{0.45\linewidth}
    \includegraphics[width=\linewidth]{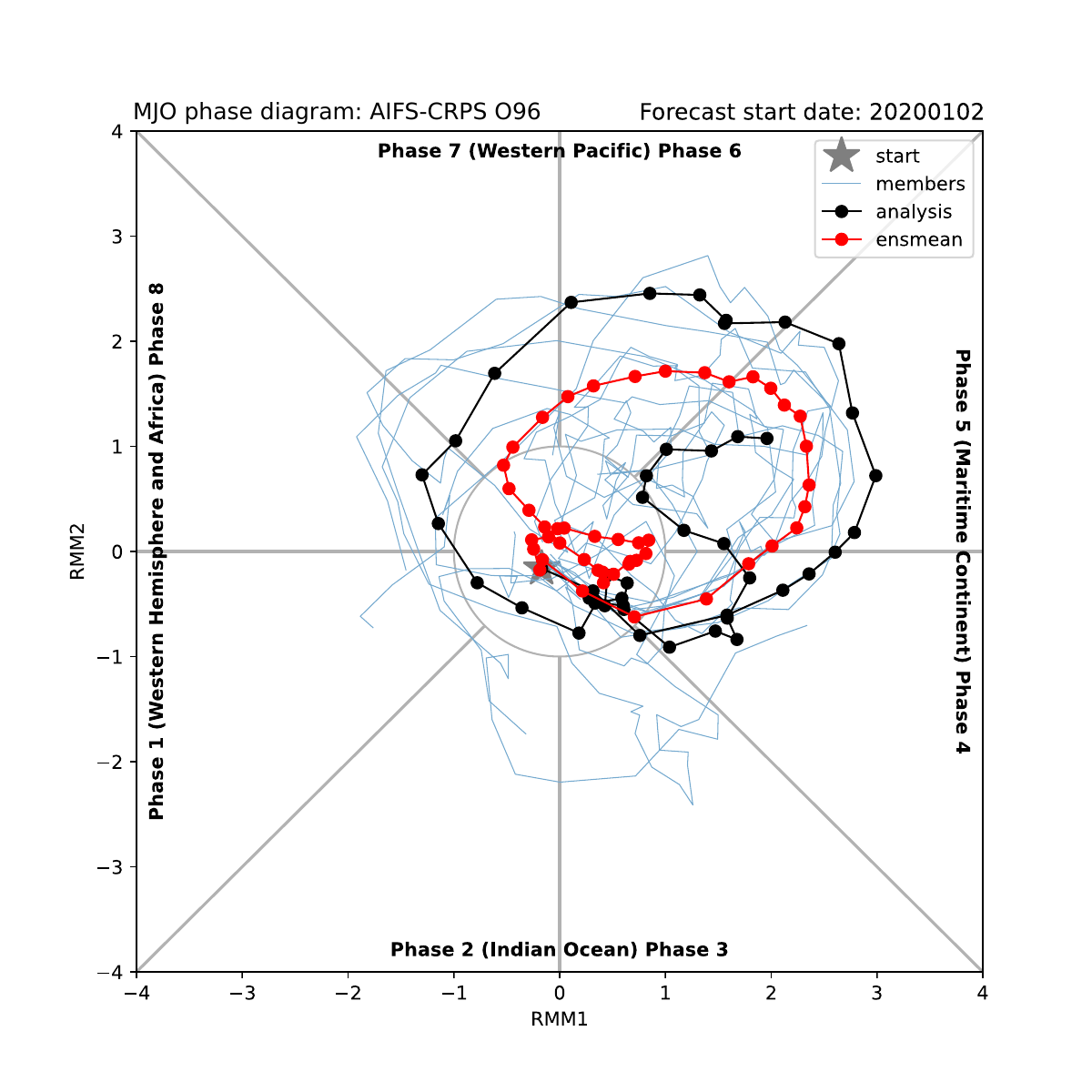}\caption{}
    \label{fig:mjo_case_aifs_phase}
\end{subfigure}
\caption{Phase diagrams based on the surrogate Real-time Multivariate MJO index described in the text for 46-day ensemble forecasts initialized on 2020-01-02 from (\subref{fig:mjo_case_ifs_phase}) IFS and (\subref{fig:mjo_case_aifs_phase}) AIFS-CRPS reforecasts.}
\label{fig:mjo_phase}
\end{figure}

To illustrate the characteristics of MJO propagation in AIFS-CRPS and IFS, we consider a case study and plot MJO phase diagrams and Hovm\"{o}ller plots (figures \ref{fig:mjo_phase} and \ref{fig:mjo_hov}) for ensemble forecasts initialized on 2020-01-02. This event is characterised by neutral MJO conditions at early lead times followed by the development of large-scale zonal wind anomalies that propagate across the Maritime Continent into the Pacific Ocean. Although our phase diagrams are constructed with a surrogate RMM index that does not include contributions from OLR, they are qualitatively extremely similar phase diagrams produced operationally by the Bureau of Meteorology (see \url{http://www.bom.gov.au/climate/mjo/}). IFS forecasts capture the development of zonal wind anomalies and initial propagation across the Maritime Continent, but they underestimate the magnitude such that the ensemble mean MJO index collapses back towards neutral conditions after $\sim$15 days. In contrast, the AIFS-CRPS forecast seems to better represent both the magnitude of the developing zonal wind anomalies and their eastward propagation across the Maritime Continent (figures \ref{fig:mjo_phase} and \ref{fig:mjo_hov}). However, we emphasize that this a single example MJO forecast and these MJO propagation characteristics may not generalise to all forecasts. 

\section{Discussion}
AIFS-CRPS shows strong medium-range forecast skill for upper-air variables. Forecasts of surface parameters are improved when resolution is increased from O96 to N320, which is consistent with deterministic forecast performance (see \cite{lang2024aifs}). At N320 resolution AIFS-CRPS shows higher forecast skill than the IFS ensemble for most surface variables.

So far, the ERA5 reanalysis and operational IFS analysis are used as input and target during training. During inference, the model is initialised with perturbed initial conditions from the IFS ensemble. Within our framework, it is possible to include perturbed initial conditions already during training. In principle, this should help the model not to alias initial condition uncertainty into model uncertainty. On the other hand, initial condition uncertainty representations, such as the ERA5 ensemble of data assimilations can be imperfect, and perturbed initial conditions can have different properties from the unperturbed reference state, which might affect auto-regressive forecasting. We will assess in future work how beneficial it is to include perturbed initial conditions during training. 

Currently, AIFS-CRPS is over-dispersive for some variables, such as geopotential at 500~hPa, in the early medium-range. This is likely related to the singular vector perturbations which are added to the initial conditions of the IFS ensemble to improve the reliability of the system. The RMSE of AIFS-CRPS is substantially lower and hence such an inflation is likely not needed. First tests with a revised initial perturbation amplitude show improved reliability (not shown). For this study, we decided to use the operational IFS initial conditions, because this is currently the most straightforward way to introduce AIFS-CRPS in an experimental real-time mode.

First tests for longer time ranges indicate that good skill can emerge from a system that has been trained on short-range forecasts only. Although our analysis of subseasonal predictability is by necessity limited to a relatively short `out-of-sample' reforecast period, the results from AIFS-CRPS are very promising. The MJO results are particularly significant as MJO forecasts from the IFS generally compare very favourably to those from other ensemble prediction systems \citep{vitart2017madden}. If these results generalise to real-time forecasts in an operational context, we expect subseasonal forecasts from AIFS-CRPS to be competitive with, or outperform, those from the best physics-based models. 

AIFS-CRPS currently shows reduced forecast skill in the stratosphere, which we believe is caused by its high sensitivity to the vertical scaling used in the afCRPS objective. We will explore revised loss definitions in future work.

\section{Conclusions}
We show that training a machine-learned weather prediction model with a proper score objective such as the afCRPS can lead to a highly skilful ensemble prediction system. AIFS-CRPS forecast skill is higher than that of the 9~km physics-based IFS medium-range ensemble for most upper-air fields and surface variables. AIFS-CRPS does not smooth the forecast fields and produces a realistic level of variability, even with long rollouts. 

Despite it being trained to optimize only short-range forecast performance, AIFS-CRPS also performs well for longer range forecasts (two to six weeks), where it exhibits lower biases and increased Madden-Julian Oscillation (MJO) forecast skill compared to ECMWF's operational subseasonal forecasting system.

AIFS-CRPS requires one single model evaluation to produce a 6~h forecast step for one ensemble member. This makes inference computationally cheap: a single member 15 day forecast is created in about one minute for AIFS-CRPS O96 and four minutes for AIFS-CRPS N320 on an NVIDIA A100 40~GB GPU, including the time spent reading the initial state and writing the forecast to disk.

It is important to note that AIFS-CRPS, like other recent probabilistic forecasting systems (e.g., \cite{price2023gencast}), relies on the analysis states of ECMWF’s physics-based NWP model for both training and forecasting. Work is ongoing at ECMWF \citep{alexe2024graphdop,mcnally2024ecmwfnl} and elsewhere \citep{vaughan2024aardvark, keller2024ai, manshausen2024generative} to explore observation-based training and initialisation, but currently these do not outperform physics-based data assimilation systems.

Future work will include exploring higher-resolution forecasts, initialise AIFS-CRPS with revised initial condition perturbations, adding more forecast parameters and testing a revised loss scaling to improve forecast skill in the stratosphere.

We expect to start running the AIFS-CRPS N320 in real-time experimental mode at ECMWF in the near future. Ensemble forecasts, meteograms and other ensemble products will be made available to the public under the terms of the ECMWF open data license.

\paragraph{Acknowledgments:} \textit{We acknowledge PRACE for awarding us access to Leonardo, CINECA, Italy. 
We acknowledge the EuroHPC Joint Undertaking for awarding this work access to the EuroHPC supercomputer MN5, hosted by BSC in Barcelona through a EuroHPC JU Special Access call. 
}

\bibliographystyle{unsrtnat}
\bibliography{refs}

\end{document}